\documentclass[12pt,manuscript=article]{achemso}
\setkeys{acs}{maxauthors = 0}
\usepackage{graphicx} % Required for inserting images
\usepackage{amsmath}
\usepackage{comment}
\usepackage[euler]{textgreek}
\usepackage[numbers]{natbib}
\usepackage{siunitx}
\usepackage{subcaption} 
\usepackage{caption}
\usepackage{longtable}
\usepackage{newfloat}
\DeclareFloatingEnvironment[name={Supplementary Table},fileext=lsst,listname={List of Supplementary Short Tables},within=section, placement=htbp!]{supptable}
\def\nlines#1{\expandafter\nlineii\romannumeral\number\number #1 000\relax}
\def\nlineii#1{\if#1m\expandafter\theline\expandafter\nlineii\fi}

\title{What to Make of Zero: Resolving the Statistical Noise from Conformational Reorganization in Alchemical Binding Free Energy Estimates with Metadynamics Sampling }

\author{Sheenam Khuttan}
\affiliation{Department of Chemistry and Biochemistry, Brooklyn College of the City University of New York, New York, NY}
\alsoaffiliation{Ph.D. Program in Biochemistry, The Graduate Center of the City University of New York, New York, NY}
\author{Emilio Gallicchio}
\email{egallicchio@brooklyn.cuny.edu}
\affiliation{Department of Chemistry and Biochemistry, Brooklyn College of the City University of New York, New York, NY}
\alsoaffiliation{Ph.D. Program in Chemistry, The Graduate Center of the City University of New York, New York, NY}
\alsoaffiliation{Ph.D. Program in Biochemistry, The Graduate Center of the City University of New York, New York, NY}

\newcommand{\sidecaption}[1]% #1 = label name
{\raisebox{\abovecaptionskip}{\begin{subfigure}[t]{1.6em}
  \caption[singlelinecheck=off]{}% do not center
  \label{#1}
\end{subfigure}}\ignorespaces}

\begin{document}

\maketitle

\begin{abstract}

We introduce the self-Relative Binding Free Energy (self-RBFE) approach to evaluate the intrinsic statistical variance of dual-topology alchemical binding free energy estimators. The self-RBFE is the relative binding free energy between a ligand and a copy of the same ligand, and its true value is zero. Nevertheless, because the two copies of the ligand move independently, the self-RBFE value produced by a finite-length simulation fluctuates and can be used to measure the variance of the model. The results of this validation provide evidence that a significant fraction of the errors observed in benchmark studies reflect the statistical fluctuations of unconverged estimates rather than the models' accuracy. Furthermore, we find that ligand reorganization is a significant contributing factor to the statistical variance of binding free energy estimates and that metadynamics-accelerated conformational sampling of torsional degrees of freedom of the ligand can drastically reduce the time to convergence.
\end{abstract}

\section{Introduction}

Advances in computational models and computer hardware are revolutionizing the role of molecular simulations in chemical research, opening new avenues for exploring molecular interactions at an unprecedented level of detail. Atomistic simulations of molecular binding, in particular, are playing a pivotal role in understanding fundamental biological regulatory processes and in assisting in the rational design of drugs \cite{abel2017advancing,armacost2020novel,zhang2021potent,Allen2022.05.23.493001,ganguly2022amber,xu2022slow}. The accurate estimation of protein-ligand binding-free energies by computer simulations, which is the subject of this study, is becoming an important ingredient in elucidating molecular recognition mechanisms, identifying potential drug candidates, and developing novel therapeutics.

However, the accurate determination of binding-free energies by physics-based atomistic computer simulations remains a formidable challenge due to the size and complexity nature of biological systems, the dynamical nature of molecular recognition mechanisms, and the high dimensionality of the conformational space to explore. The presence of many energy basins separated by high energy barriers is a serious obstacle for traditional molecular dynamics (MD) conformational sampling algorithms, which are limited to the narrow band of thermal energies. The negative impact of conformational trapping due to limited MD conformational sampling is further exacerbated in simulations of molecular association processes where the populations of conformational states of the receptor and ligand often shift as they form interactions \cite{Mobley2012,procacci2019solvation}. Poor equilibration between stable configurations of the system and failure to capture the free energy of conformational reorganization upon the formation of receptor-ligand interactions causes biased and noisy free energy estimates that do not reflect the actual binding affinity trends, leading to incorrect predictions about the relative potency of drug candidates.

Alchemical models of the Relative Binding Free Energies (RBFE) of protein-ligand complexes have emerged as the leading computational methods for lead optimization in industrial and academic pharmaceutical research \cite{GallicchioSAMPL4,wang2015accurate,zou2019blinded,schindler2020large,lee2020alchemical,kuhn2020assessment,gapsys2020large,bieniek2021ties,hahn2022bestpractices,gapsys2022pre,sabanes2023validation,cournia2017relative}.  RBFE models estimate the ratio of the dissociation constants, $K'_d/K_d$, of a pair of ligands to the same protein receptor, or, equivalently, their relative standard binding free energies, $\Delta\Delta G^\circ_b$ by considering a non-physical path that progressively modifies the potential energy function of the system in such a way that at the beginning it describes the receptor bound to the first ligand and at the end it describes the receptor bound to the other ligand. The relative binding free energy is then the reversible work along the alchemical path \cite{Jorgensen2004,cournia2017relative,Mey2020Best,azimi2022relative}. 

While increasingly popular, as evidenced by extensive large-scale benchmarking validation studies against experimental data\cite{wang2015accurate,schindler2020large,gapsys2020large,hahn2022bestpractices,sabanes2023validation}, RBFE models do not always yield correct predictions. The causes of mispredictions are often unclear; primarily because the ground truth value of the models is not known, and the relative contributions of model accuracy and statistical fluctuations on the prediction accuracy are uncertain. Are prediction errors caused by inaccuracies in the models or our inability to calculate the models' predictions with sufficient precision? In this work, we investigate the causes of slow convergence and large statistical fluctuations of relative binding free energy estimates on a large and challenging library of protein-ligand complexes. We do so by investigating calculations that connect chemically identical complexes and should then yield zero. We can then explore the models' bias and variance independently. We find that the conformational reorganization of the ligand is a leading cause of poor convergence and that an accelerated conformational sampling approach based on metadynamics can significantly reduce statistical fluctuations.     

Alchemical RBFE models are still a work in progress as structure-based drug discovery aids in many respects. RBFE tools tend to be very complex, require extensive expertise, and display inconsistent performance if not deployed correctly.\cite{schindler2020large} Probably some fraction of the RBFE prediction errors that are observed in applications are caused by erroneous chemical representations, such as the incorrect assignment of protonation, tautomerization, and chirality. Inaccuracies of molecular mechanics potential energy functions are also likely a significant source of errors \cite{liu2013lead,Mey2020Best}. Technical difficulties exist in many alchemical RBFE implementations with charge-changing and scaffold-hopping transformations, and in modeling variations of hydration patterns. However, as discussed above, limited conformational sampling likely remains one of the primary sources of mis-predictions. The system often stays trapped near the initial conformation, and alternative poses of the receptor and the ligands, including the conformational reorganization processes occurring upon binding, are not fully captured during the relatively short molecular dynamics runs.\cite{sakae2020absolute,khuttan2023taming}

There are many alchemical RBFE implementations in current use. The Double Decoupling Method (DDM) \cite{Gilson:Given:Bush:McCammon:97}, which is probably the most popular, does not connect the end states directly. Rather it relies on an indirect route involving multiple simulations that morph the electrostatic and non-electrostatic interaction of one ligand into the other in the solution and receptor environments separately \cite{rocklin2013calculating}. The implementation of DDM typically requires customized MD energy routines that allow the tuning of the parameters of the potential energy function as the alchemical transformation takes place and incorporate modified soft-core interaction pair-potentials to reduce numerical instabilities near the endpoints. 

The treatment of the transformation of the chemical topology of one ligand into the other is an important differentiating factor of alchemical RBFE implementations. In a single-topology implementation, the system holds a single representation of the ligands' atoms and their assigned force field parameters in such a way that the atoms of the initial molecule are converted into those of the final molecule during the alchemical transformation. Dummy atoms are used to treat atoms that are not present at either end state \cite{fleck2021dummy,zou2019blinded,jiang2019computing,Gallicchio2021binding}. Conversely, in a dual-topology RBFE implementation, the two ligands are represented by distinct non-interacting standard chemical topologies whose interactions with the environment are turned off and on during the alchemical process \cite{azimi2022relative}. Hybrid topologies, where the constant parts of the ligands are treated within the single-topology formalism and the variable parts are described by dual-topology, are also in use. Single- and dual-topology approaches are more or less suitable depending on the circumstances. Generally, single-topology RBFE is more efficient, especially when the difference between the two ligands is small, and dual-topology RBFE formulations are more versatile and easier to implement. 

We recently introduced the Alchemical Transfer Method (ATM) to address some of the complexities and limitations of traditional alchemical methods. ATM is a dual-topology RBFE implementation based on a coordinate rather than a potential energy function perturbation. ATM is free of the complexities of traditional alchemical methods. It supports absolute and relative binding free energy calculations in a unified way, it computes free energies directly employing a single simulation box with standard chemical topologies, and it natively supports standard as well as charge-changing and scaffold-hopping transformations without correction factors and ancillary calculations. Furthermore, since it does not use parameter interpolation or custom soft-core pair potentials, ATM is more easily implemented and transferable across MD engines because it uses the unmodified energy routines of the underlying molecular dynamics engine. For the same reason, it applies to any molecular energy function, including the next generation of more accurate polarizable,\cite{harger2017tinker,panel2018accurate,Huang2018drude,das2022development} quantum-mechanical,\cite{beierlein2011simple,lodola2012increasing,hudson2019use,casalino2020catalytic} and machine-learning potentials\cite{smith2019approaching,rufa2020towards,eastman2023openmm} that are just starting to be employed in alchemical macromolecular simulations. The current fully open-source software release of ATM employs the OpenMM molecular dynamics engine and has been successfully tested on a series of medicinal targets by us and academic and industrial partners.\cite{sabanes2023validation,chen2023performance} 

In this work, we study the bias and variance of ATM by estimating the binding free energies of a series of complexes from the benchmark set of Schindler et al.\cite{schindler2020large} relative to themselves (self-RBFEs).  A self-RBFE is obtained when the two ligands considered in an ATM RBFE calculation are the same ligand. Obviously, in this case, the true value of the RBFE is zero. Nevertheless, because the dual-topology copies of the ligand act independently, the free energy value produced by a finite-length ATM simulation fluctuates and is not guaranteed to be zero. The advantage of investigating self-RBFEs is that their true value (zero) is known, allowing the bias and variance of the model to be investigated independently. We measure the bias by asking how much the average of a sequence of the ATM self-RBFE replicates differs from zero. The variance is then measured from the distribution of the replicate's estimates. While a large and consistent bias reflects an implementation error that should be corrected, a level of variance is unavoidable and reflects the minimum amount of statistical noise that would affect actual ATM's RBFE predictions between pairs of different ligands.

We observe that ATM's self-RBFE variance is a significant fraction of the mean squared error of ATM RBFE estimates relative to experimental free energies in recent large-scale validation studies,\cite{sabanes2023validation,chen2023performance} suggesting that, to some degree, those errors reflect statistical noise rather than model's defects that can be addressed by improving the chemical realism of the model by, for example, adopting a potential energy model at a higher level of theory. Rather, improved predictions could be achieved by reducing statistical noise by more extensive conformational sampling. This conclusion is supported by the observation that the self-RBFE's variance is strongly correlated to the reorganization free energy of the ligand; the induced-fit free energy cost for the ligand to reorganize into the binding-competent conformational state from the range of conformations it occupies in solution. A link between these two quantities suggests that ligand conformational reorganization contributes significantly to the errors observed in validation studies and that errors can be reduced by improving the sampling of the ligands' intramolecular degrees of freedom. 

In this work, we employ metadynamics-based sampling\cite{barducci2008well} to speed up the sampling of slow torsional degrees of freedom during ATM RBFE calculations. Metadynamics is an algorithm that adaptively builds up a biasing potential function that disfavors conformations that have already been visited. By doing so, it tends to equalize the populations of conformational states along a chosen coordinate and lower energy barriers that hamper rapid interconversions. In keeping with the philosophy of simplicity and transferability of ATM, we employ the metadynamics implementation in OpenMM by Peter Eastman\cite{eastman2023openmm} that, unlike other conformational acceleration algorithms such as replica exchange with solute tempering (REST),\cite{wang2013modeling} apply to arbitrary many-body potential and does not require modifications to the core energy routines of the MD engine.

The work is organized as follows, we first review ATM, then introduce the concepts of self-RBFE and reorganization free energies, and describe the metadynamics algorithm as used in this study. We then present the analysis of the self-RBFE and reorganization free energy values we obtain on the benchmark sets. We conclude with a discussion of the implications of the findings of this study for the future directions of alchemical binding free energy models in structure-based drug discovery.

\section{Theory and Methods}

\subsection{The Alchemical Transfer Method for Relative Binding Free Energy Estimation}

The Alchemical Transfer Method (ATM, for short) estimates the binding free energies of molecular complexes by relating the bound and unbound states by a coordinate displacement transformation that brings the ligand from the solution environment to the binding site of the receptor. Alternatively, it estimates the relative binding free energy (RBFE) of two complexes of the same receptor with two different ligands by translating one ligand into the binding site while another is simultaneously translated from the binding site to the solution. In this sense, ATM is a dual-topology free energy method because it employs distinct topologies for each ligand rather than modifying one topology as in single-topology formulations.\cite{Mey2020Best} ATM and its applications are described in detail in published works.\cite{wu2021alchemical,azimi2022relative} Only a brief account is provided here to introduce the notation and the essential features relevant to the present work.

\begin{figure*}
    \centering
    \includegraphics[scale=0.65]{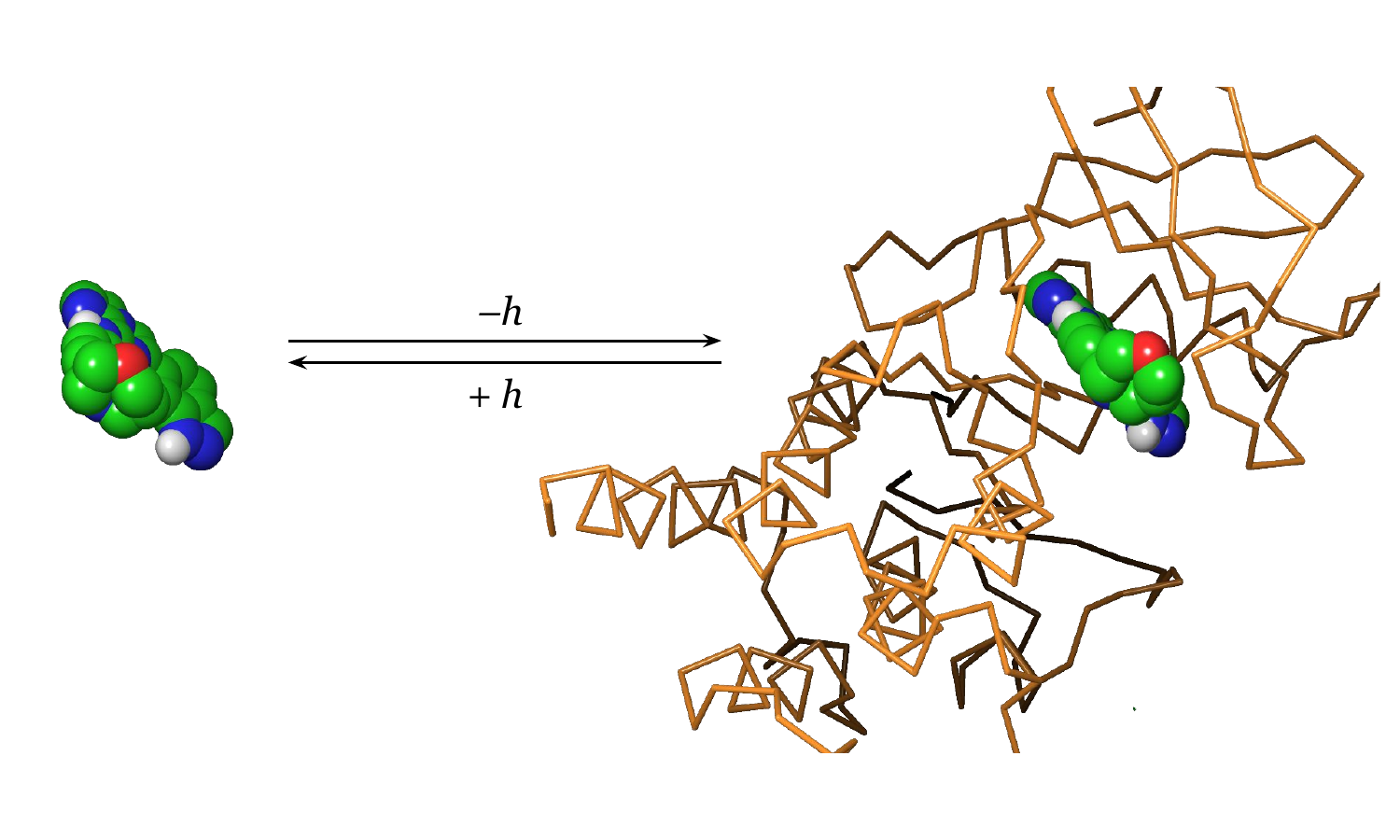} 
    \caption{\label{fig:atm-transfer} General illustration of the Alchemical Transfer Method (ATM) for self-RBFE. The unbound guest is obtained by translating the bound guest by the displacement vector $h$ shown in black. The direction of $h$ depends on the translation towards and from the binding site. The protein backbone of the Syk protein in orange is shown along with the ligand CHEMBL3265036 from the Schindler et al.\cite{schindler2020large} protein-ligand database. The green, blue, red and white atoms of the ligand represent carbon, nitrogen, oxygen, and hydrogen atoms, respectively.
 }   
\end{figure*}

A typical system for an ATM RBFE calculation consists of a protein receptor ${\rm R}$ bound to a ligand $\rm A$ and a second ligand $\rm B$ placed in the solvent displaced from ligand $\rm A$ by a displacement vector $h$, such that it is at a sufficient distance from the receptor to be considered not bound to it (Figure \ref{fig:atm-transfer}). ATM computes the potential energy functions of the system and their gradients before and after translating by a vector $h$ ligand $\rm A$ from the binding site to the solvent while simultaneously translating ligand $B$ by the opposite displacement. The first potential energy function, called $U_0(x)$, describes the system when ligand $\rm A$ is bound to the receptor, and the second, called $U_1(x)$, corresponds to the state in which the ligand $\rm B$ is bound. Here, $x$ represents collectively the coordinates of the receptor, the ligands, the solvent, and whatever other chemical species is present in the system. The free energy difference between the states $1$ and $0$ is the RBFE between ligands $\rm A$ and $\rm B$ to receptor $\rm R$.

To calculate the RBFE, the potential energy function is progressively morphed from $U_0(x)$ to $U_1(x)$ by defining an alchemical potential energy function $U_\lambda(x)$ that goes from $U_0(x)$ to $U_1(x)$ as the alchemical progress parameter $\lambda$ goes from $0$ to $1$. As an example, the linear alchemical potential energy function
\begin{equation}
U_\lambda(x) = U_0(x) + \lambda u(x) \, ,
\label{eq:Ulambda-linear-def}
\end{equation}
where
\begin{equation}
u(x) = U_1(x) - U_0(x) \, ,
\label{eq:upert-linear-def}
\end{equation}
is the {\em perturbation energy function}, is one such interpolating function. However, as thoroughly discussed in published works,\cite{pal2019perturbation,khuttan2021alchemical} non-linear alchemical potential energy functions are vastly more efficient than linear interpolating functions. ATM adopts the expression 
\begin{equation}
U_\lambda(x) = U_0(x) + W_\lambda[u(x)] \, ,
\label{eq:Ulambda-nonlinear-def}
\end{equation}
where $W_\lambda(u)$ is the soft-core softplus alchemical perturbation function 
\begin{equation}
  W_{\lambda}(u)=\frac{\lambda_{2}-\lambda_{1}}{\alpha}\ln\left\{1+e^{-\alpha [u_{\rm sc}(u)-u_{0}]}\right\}+\lambda_{2}u_{\rm sc}(u)+w_{0} .\, ,
  \label{eq:softplus-function}
\end{equation}
the parameters $\lambda_{2}$, $\lambda_{1}$, $\alpha$, $u_{0}$, and $w_{0}$ are functions of $\lambda$ (see Computational Details),
the function
\begin{equation}
  u_{\rm sc}(u)=
\begin{cases}
u & u \le u_c \\
(u_{\rm max} - u_c ) f_{\rm sc}\left[\frac{u-u_c}{u_{\rm max}-u_c}\right] + u_c & u > u_c
\end{cases}
\label{eq:soft-core-general}
\end{equation}
with
\begin{equation}
f_{\rm sc}(y) = \frac{z(y)^{a}-1}{z(y)^{a}+1} \label{eq:rat-sc} \, ,
\end{equation}
and
\begin{equation}
    z(y)=1+2 y/a + 2 (y/a)^2
\end{equation}
is the soft-core perturbation energy function designed to avoid singularities near the initial state of the alchemical transformation.\cite{pal2019perturbation,khuttan2021alchemical} The parameters $u_{\rm max}$, $u_c$, and $a$ are set to cap the perturbation energy $u(x)$ to a maximum positive value without affecting it away from the singularity. The specific values of $u_c$, $u_{\rm max}$, and of the scaling parameter $a$ used in this work are listed in the Computational Details.

For efficiency reasons elaborated elsewhere,\cite{wu2021alchemical,azimi2022relative} Eq.\ (\ref{eq:Ulambda-nonlinear-def}) is not employed to span the entire alchemical pathway from $\lambda = 0$ to $\lambda=1$. Rather, the process is divided into two legs: one starting at $\lambda = 0$ using the alchemical potential in Eq.\ (\ref{eq:Ulambda-nonlinear-def}), and a second leg starting from the bound state $U_1(x)$ morphing in the other direction towards the unbound state using the alchemical potential $U_\lambda(x) = U_1(x) + W_{1-\lambda}[-u(x)]$. Both legs terminate at $\lambda=1/2$ at the ATM symmetric alchemical intermediate with the potential energy function $U_{1/2}(x) = [U_0(x)+U_1(x)]/2$ that is an equally weighted average of the endstates. The relative binding free energy is then given by the differences of the free energies corresponding to the two legs
\begin{equation}
    \Delta \Delta G_b = \Delta G_b(B) - \Delta G_b(A) = \Delta G_{\rm leg1} - \Delta G_{\rm leg2}
\end{equation}

\subsection{Self-Relative Binding Free Energy Calculations}

In this work, we investigate self-RBFE ATM estimates; that is the outcomes of RBFE calculations when the ligands $A$ and $B$ are the same ligand. The true value of the binding free energy of a ligand relative to itself is obviously zero. However, because of statistical fluctuations, the self-RBFE obtained from a finite-length ATM calculation will not be exactly zero. Below we will employ the statistical fluctuations of self-RBFE estimates to understand the statistical fluctuations of RBFEs between unlike ligands.

It should be recognized that the concept of a self-RBFE applies only to dual-topology binding free energy formulations such as ATM. The single-topology process of morphing a ligand to the same ligand is inherently a null transformation with necessarily zero free energy.  In ATM theory, the true value of a self-RBFE is zero because the two legs of the ATM alchemical process have the same initial and final states and thus have the same free energy. The initial state of either leg is the state in which one copy of the ligand is bound to the receptor, and the other copy is in solution. The final state is the symmetric alchemical intermediate, which is again shared by the two legs. However, the free energy of each leg is not zero, and random differences between the estimates of the two legs cause the self-RBFE estimate to differ from zero.

We measure the statistical fluctuations of self-RBFEs for a set simulation length by running replicates of the simulations of the same length. The standard deviation of the distribution of self-RBFEs is a measure of the statistical fluctuation of the self-RBFE of a ligand. The deviation of the mean of the self-RBFE of the distribution of self-RBFEs from zero is a measure of the bias of the ATM estimate. 

\subsection{Estimation of the Ligand Binding Reorganization Free Energy}

% most of this text belongs to the intro or discussion. We'll keep it here for now.

The ligand reorganization free energy for binding measures the free energy cost for the ligand in solution to assume the binding-competent conformation.\cite{Gallicchio2011adv} This quantity, also known in the literature as the conformational free energy penalty or strain energy,\cite{Foloppe2016TowardsUT,Foloppe2021TheRE} is an important element considered in lead optimization because a molecule predisposed for binding with small reorganization free energy is more likely to bind strongly to the receptor. Conversely, reorganization opposes the binding of flexible molecules that spend most of their time in solution in conformations away from the bioactive conformation.
Even though drug development typically focuses on strengthening receptor-ligand interactions, the ligand reorganization element can be crucial in determining binding specificity,\cite{Yang2009ImportanceOL} especially when binding energy variations are minimal. In such cases, optimizing binding affinity can be achieved by strategies that focus on preorganizing the ligand for binding, thereby reducing unfavorable reorganization. 

Computer models are uniquely positioned to probe ligand reorganization free energies. While experimental structures of protein-ligand complexes often yield the bound structure of ligands, they do not provide information about their distribution of conformations in solution. The ligand component of the reorganization free energy for binding of ligand $\rm A$, $\Delta G_{\rm reorg}(A)$, is formally related to the population $p_A$ of the ligand's bioactive conformation in solution
\begin{equation}
    \Delta G_{\rm reorg}(A) = -k_B T \ln p_A \, .
\end{equation}
Hence, the binding affinity of a ligand could be significantly overestimated if the reorganization free energy is not taken into account, especially if $p_A$ is small and $\Delta G_{\rm reorg}(A)$ is large and positive.

It is challenging to obtain a good estimate of the reorganization free energy of a ligand by unbiased molecular simulations if the bioactive conformation is rarely visited in solution or it is separated by the other solution conformations by large energy barriers that are rarely crossed. For example, if the ligand remains trapped in the initial bound conformation because transitions to more stable conformations in solution are rare, one would incorrectly deduce that the ligand reorganization free energy is small.  The equilibration between conformational states separated by free energy barriers greater than 5 kcal/mol is generally considered difficult within routine MD simulation timescales. Conversely, the reorganization free energy could be grossly overestimated if the population of the bioactive conformation is so small that is almost never visited when the ligand is in solution. The reorganization of the receptor can also be a major contributing factor for binding that is equally or harder to model than ligand reorganization. In this work, we focus on the ligand reorganization under the assumption that the receptor does not reorganize or it does not reorganize differently depending on the bound ligand. 

Taken together, conformational reorganization processes that accompany binding can be a major convergence bottleneck for binding free energy calculations. One way to probe their effect is to see whether ligands with self-RBFEs with large statistical fluctuations are predominantly those that reorganize upon binding. This would be the case because ligand molecules that suffer large reorganization in solution and rarely interconvert between the bound and unbound conformations in solution are more likely to display random fluctuations in one alchemical leg than the other, causing self-RBFEs to deviate from zero.

In this work, we estimate the reorganization free energy of a ligand $A$ by measuring the RBFE between the ligand and a version of the same ligand restrained within the state corresponding to the bioactive conformation $A_b$. The bound state is identified by means of ranges of torsional angles of the ligand. According to the two-step process
\begin{align}
 \label{scheme:reorg} 
R + A   &\rightleftharpoons  R + A_b   & \Delta G_{\rm reorg}(A) \\ \nonumber
R + A_b &\rightleftharpoons  RA_b      & \Delta G_b(R,A_b) \nonumber
\end{align}
where $A_b$ is the restrained ligand pre-reorganized for binding, $\Delta G_{\rm reorg}(A)$ is the reorganization free energy, $\Delta G_b(RA_b)$ is the binding free energy of the reorganized ligand, and assuming that the complexes $RA_b$ and $RA$ with and without the ligand restrained are thermodynamically equivalent, the overall binding free energy for the process $R + A \rightleftharpoons RA $ is written as\cite{Gallicchio2011adv}
\begin{equation}
    \Delta G_b(R,A) =  \Delta G_{\rm reorg}(A) + \Delta G_b(R,A_b)
\end{equation}
Thus, according to this model, the reorganization free energy is given by the relative binding free energy between $A$ and $A_b$ for receptor $R$.

\begin{figure*}
    \centering
    \includegraphics[scale=1.0]{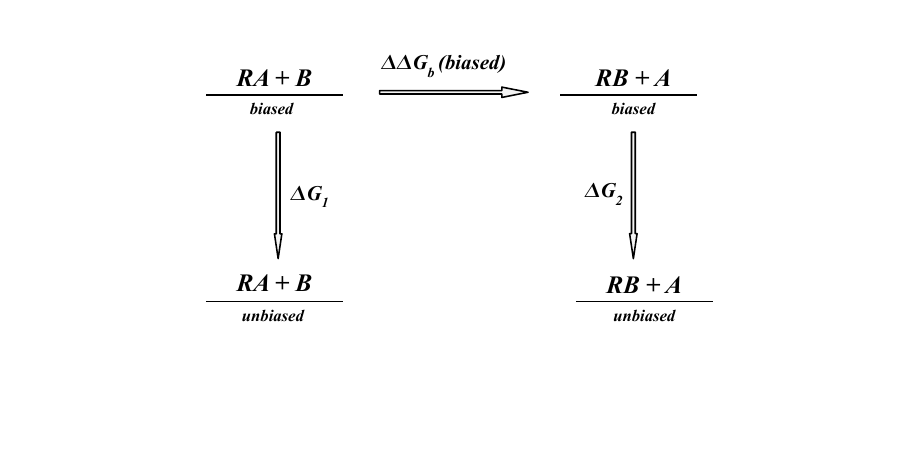} 
    \caption{ \label{fig:atm-metaD} Thermodynamic cycle for the unbiasing of the relative binding free energy between ligands $A$ and $B$ calculation of the unbiased $\Delta\Delta G_b$ in the ATM-metaD protocol. In the self-RBFE approach, both ligands A and B are chemically identical and are bound to the receptor R at the two end states. }
   
\end{figure*}

\subsection{Metadynamics Conformational Sampling}

In this study, we employ well-tempered metadynamics\cite{barducci2008well} to accelerate the sampling of the internal torsional degrees of freedom of the ligands. Similar to earlier local elevation\cite{huber1994local} and conformational flooding methods,\cite{grubmuller1995predicting} metadynamics constructs a flattening biasing potential that reduces free energy barriers along chosen collective variables (CVs) by disfavoring values of the CVs that are frequently visited.\cite{laio2002escaping,iannuzzi2003efficient} At the limit at which free energy barriers are completely removed, well-tempered metadynamics biasing potential yields the potential of mean force of the system along the selected CVs.\cite{barducci2008well} The results reported here were obtained with the implementation of well-tempered metadynamics by Peter Eastman packaged with the OpenMM library.\cite{eastman2017openmm,OpenMM-Metadynamics} 

As described in detail in Computational Details, our approach involves first obtaining the torsional flattening biasing potentials $U_{\rm bias}$ by running well-tempered metadynamics on the free ligands in water. We then perform the ATM alchemical calculation with the biasing potentials added to the intramolecular potential energy functions of the ligands. The resulting biased free energy is unbiased using a book-ending approach\cite{hudson2019use,khuttan2023taming} by computing the free energy differences of the system without the biasing potential from samples collected with the biasing potential at the endpoints of the alchemical path. In this work, we used a simple unidirectional exponential averaging formula to evaluate the free energy corrections for unbiasing at each endpoint (Figure \ref{fig:atm-metaD}). For example, in the notation of Figure \ref{fig:atm-metaD}
\begin{equation}
\Delta G_1 = -k_B T \ln \langle \exp(U_{\rm bias}/k_B T) \rangle_{\rm biased}
\end{equation}
where $U_{\rm bias}$ is the metadynamics-derived flattening biasing potential and $\langle \ldots \rangle_{\rm biased}$ is the ensemble average of the $RA + B$ biased state. The exponential averaging estimator converges quickly in this case because the biased ensemble is a subset of the unbiased one. 

The ATM protocol augmented with metadynamics (hereafter ATM+MetaD) was applied to the calculation of ligand reorganization free energies (see above) and to the self-RBFE calculations. As shown in the Results, metadynamics sampling reduces significantly the statistical fluctuations of the self-RBFE estimates by accelerating the transitions of the ligands. 

\subsection{Benchmark Systems}

\begin{figure*}
    \centering
    \includegraphics[scale=0.50]{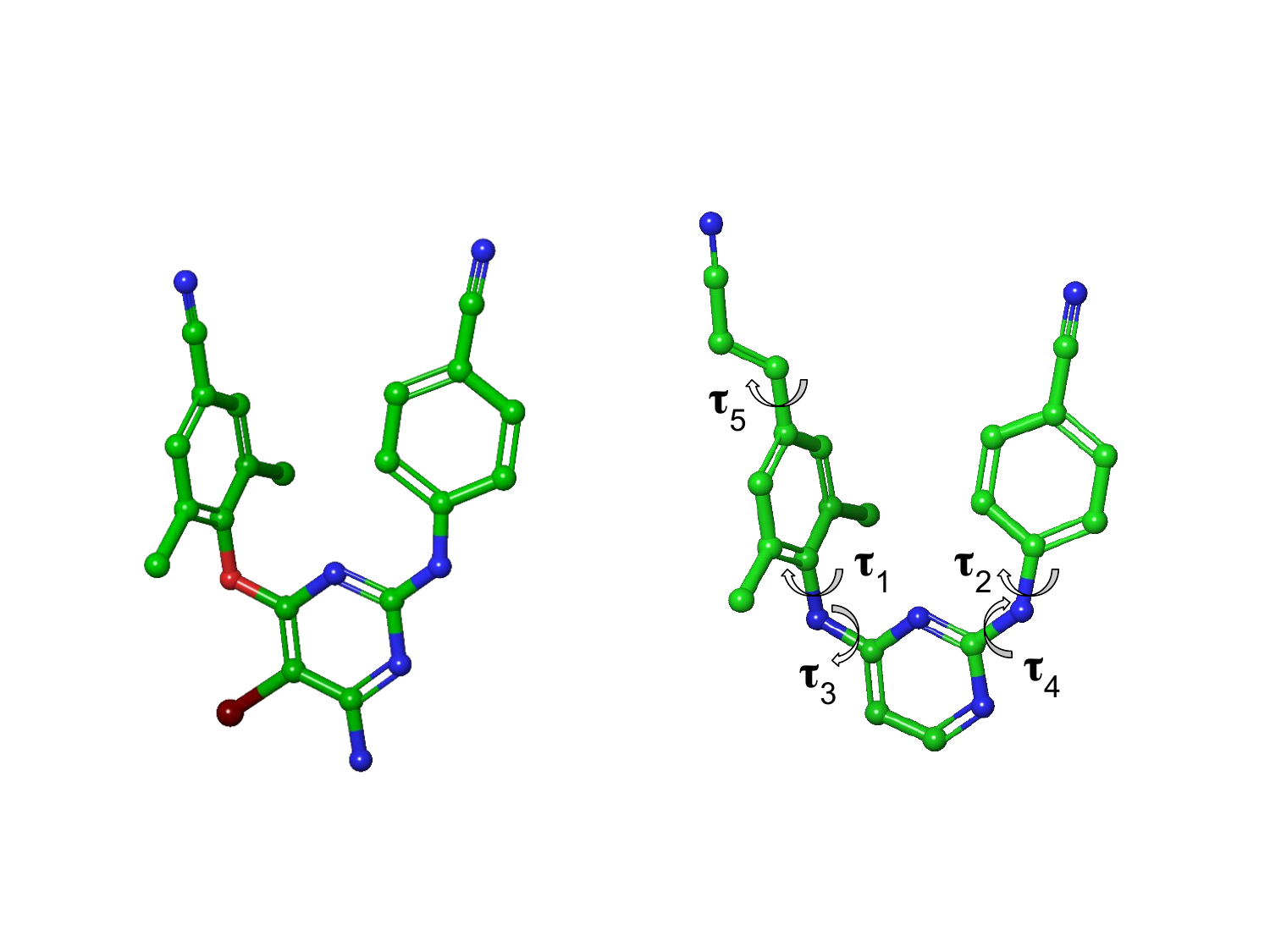} 
    \caption{\label{fig:nnrtis} Chemical structures of TMC125 (etravirine)  TMC278 (rilpivirine). Hydrogen atoms are not shown. The nomenclature of the torsional angles is indicated for TMC278. Carbon atoms are depicted in green color, nitrogen atoms in blue, oxygen atoms in red, and the bromine atom in maroon. The same nomenclature applies to TMC125, except for the $\tau 5$ angle, which is specific to TMC278. }   
\end{figure*}

We illustrate the application of the self-RBFE and reorganization free energy analysis on four benchmark systems. The first set, composed of two non-nucleoside inhibitors (NNRTIs) (TMC125 and TMC278) of HIV-1 reverse transcriptase (HIV-RT) whose conformational reorganization properties have been studied computationally (Figure \ref{fig:nnrtis}),\cite{Okumura2010,Gallicchio2012b} has been chosen as one of the clearest examples of conformational reorganization. TMC278, sold as rilpivirine, is known to undergo an extensive conformational reorganization from an extended conformation to a compact U-shaped conformation to bind HIV-RT. In contrast, the similar TMC125 compound (sold as etravirine) is mostly in the binding-competent U-shaped conformation in solution and does not reorganize for binding. 

The other three case study systems (c-Met, Syk, and CDK8) were taken from the RBFE benchmark set prepared by Schindler et al.\cite{schindler2020large} The RBFEs estimates of these systems were recently reported by Chen et al.\cite{chen2023performance} using the same ATM free energy protocol and force field employed here. Chen et al.\cite{chen2023performance} reported average unsigned errors relative to the experiments of $0.98$, $1.13$, and $1.50$ kcal/mol, respectively, relative to the experiments for these sets. Schindler et al.\cite{schindler2020large} obtained similar prediction performance with the commercial FEP+ package.\cite{wang2015accurate,wang2017accurate}
Here we report the self-RBFE estimates for all of the 101 complexes in these sets and perform a detailed statistical fluctuation and reorganization free energy analysis on a randomly picked subset of 11 complexes.

\subsection{Simulation Settings}

We employed the structures of the c-Met, Syk, and CDK8 complexes posted by Schindler et al.\cite{schindler2020large,SchindlerRepository} prepared and parameterized for ATM RBFE calculations as described by Chen et al.\cite{chen2023performance} and posted on the GitHub repository {\tt https://github.com/EricChen521/ATM\_MerckSet}.

We used the 3MEC\cite{Lansdon2010} and 2ZD1\cite{Das2008} crystal structures for the complexes of TMC125 and TMC278 bound to HIV-RT. The HIV-RT receptor structures were processed using the Protein Preparation Wizard in Maestro (Schr\"{o}dinger, Inc).  Given the large size of the HIV-RT enzyme complex, the calculations used a simplified model of the receptor, incorporating any receptor residue with atoms within 12 Å of any ligand atom. The 12 Å limit for non-bonded interactions and the receptor's rigidity ensure that predictions do not significantly depend on atomic interactions outside this modeled zone. The receptor model consisted of 114 residues (from positions 88 - 112, 171 - 195, 220 - 243, 314 - 323, 347 - 350, and 378 - 385 in the \textit{p66} subunit, and 132 - 142 in the \textit{p51} subunit), totaling 1905 atoms. We adopted the AMBER FF14SB force field for the protein and the TIP3P model for water. TMC125 and TMC278 were parametrized using the GAFF1.8/AM1-BCC forcefield. The solvated systems were neutralized using Na$^+$/Cl$^-$ and K$^+$/Cl$^-$ ions for HIV-RT protein-ligand and Schindler et. al benchmark sets respectively. ATM ligand alignment restraints\cite{azimi2022relative} were employed with force constants $k_r = 2.5$ kcal/(mol \AA$^2$) and $k_\theta = k_\psi = 25.0$ kcal/mol for the positional and orientational restraints, respectively. The receptors' C$\alpha$ atom positions were kept near their starting values using flat-bottom harmonic restraints with a $1.5$ \AA\ allowance and a force constant of $25$ kcal/(mol \AA$^2$). The LEaP program\cite{AmberTools} was used to combine the receptor and the ligands and solvate the system. The second ligand of each ligand pair was translated by $34$ \AA\ along the diagonal of the solvent box. This distance was sufficient to maintain a minimum separation of three water layers between the ligand in solution and the receptor's atoms. The resulting system was solvated with a $10$ \AA\ buffer. 

The prepared systems were energy-minimized, thermalized, and equilibrated at 300 K and 1 bar of constant pressure. This was followed by slow annealing to the $\lambda = 1/2$ alchemical intermediate for 250 ps. The resulting structure served as the initial configuration for the subsequent alchemical replica exchange simulations.

For all the protein-ligand complexes, we employed 22 replicas, with 11 for each of the two legs from $\lambda=0$ to $\lambda=1/2$. The schedules of the softplus alchemical parameters in Equation \ref{eq:softplus-function} were: $\lambda_1 = 0$, $0$, $0$, $0$, $0$, $0$, $0.1$, $0.2$, $0.3$, $0.4$, $0.5$, $\lambda_2 = 0$, $0.1$, $0.2$, $0.3$, $0.4$, $0.5$, $0.5$, $0.5$, $0.5$, $0.5$, $0.5$ for the Schindler et. al. protein-ligand systems. For HIV-RT complexes, the parameters were $\lambda_1 = 0$, $0$, $0$, $0$, $0$, $0$, $0.10$, $0.20$, $0.30$, $0.40$, $0.50$, $\lambda_2 = 0$, $0.10$, $0.20$, $0.30$, $0.40$, $0.50$, $0.50$, $0.50$, $0.50$, $0.50$, $0.50$. For all sets and $\lambda$-states we set $\alpha = 0.1$ (kcal/mol)$^{-1}$, and $u_0 = 110$ kcal/mol. We used the softcore perturbation energy parameters $u_{\rm max} = 200$ kcal/mol, $u_c = 100$ kcal/mol, and $a = 1/16$. 

The asynchronous Hamiltonian replica exchange molecular dynamics conformational sampling \cite{gallicchio2015asynchronous}, was executed with a timestep of 2 fs. Perturbation energy samples were collected every 40 ps. The relative binding free energies were determined using replica trajectories that were a minimum of 5 ns in length. The first third of the samples were discarded for equilibration for free energy analysis. We used UWHAM, multi-state free energy estimator, \cite{Tan2012} for free energy, and statistical error estimation.

We employed the well-tempered metadynamics\cite{barducci2008well} utility available in OpenMM\cite{eastman2023openmm} to optionally accelerate the sampling of the torsional degrees of the ligands. Torsional potential energy flattening biasing potentials were obtained by simulating each ligand in a water solution. Metadynamics MD was conducted for 50 ns with a well-tempered metadynamics bias factor of 9, Gaussian height $0.3$ kJ/mol, $10$ degrees Gaussian width, and a deposition rate of  0.2 ps. The biasing potential was added to the potential energy function of the ligands for a subset of the self-RBFE ATM calculations, and the resulting free energy estimates were unbiased, as described in Theory and Methods.

For the calculation of ligand reorganization energy, a set of torsions of one copy of the ligand was restrained to keep it in the bound state. To establish the range of allowed torsional angles, we obtained the distribution of torsional angles of the ligand when bound to the receptor ($\lambda = 1$) during the self-RBFE calculation. The tolerances of the torsional restraints were set so as to cover most of the span of the distributions. The specifications for the torsional restraints for each ligand in this study can be found at ({\tt https://github.com/sheenam1509/self-rbfe.git}). The second copy of the ligand was unrestrained and was modeled with the metadynamics-derived biasing potential described above to avoid conformational trapping. 

\section{Results}

The self-RBFE estimates from five replicates for the complexes of HIV-RT with TMC125 and TMC278 with and without accelerated metadynamics sampling are reported in Table \ref{tab:replicates-hiv}. The average and variance over the five replicates measure the bias and statistical fluctuations of the ATM self-RBFE estimator for a simulation length of 5 ns per replica. With the exception of the complex with TMC278 with metadynamics sampling, the bias of the self-RBFE with respect to the true value is zero within statistical uncertainty. However, a small but consistent bias towards positive self-RBFE values is evident as the averages over the replicates are positive in all cases, and the average self-RBFE is outside the uncertainty window in the case of TMC278 with metadynamics. We interpret this residual bias as the tendency of the copy of the ligand started in solution to move away from the bound-competent conformation in the early phases of the simulation. 

\begin{figure}
  \centering
  \sidecaption{subfig:a1}
  \raisebox{-\height}{\includegraphics[width=0.70\textwidth]{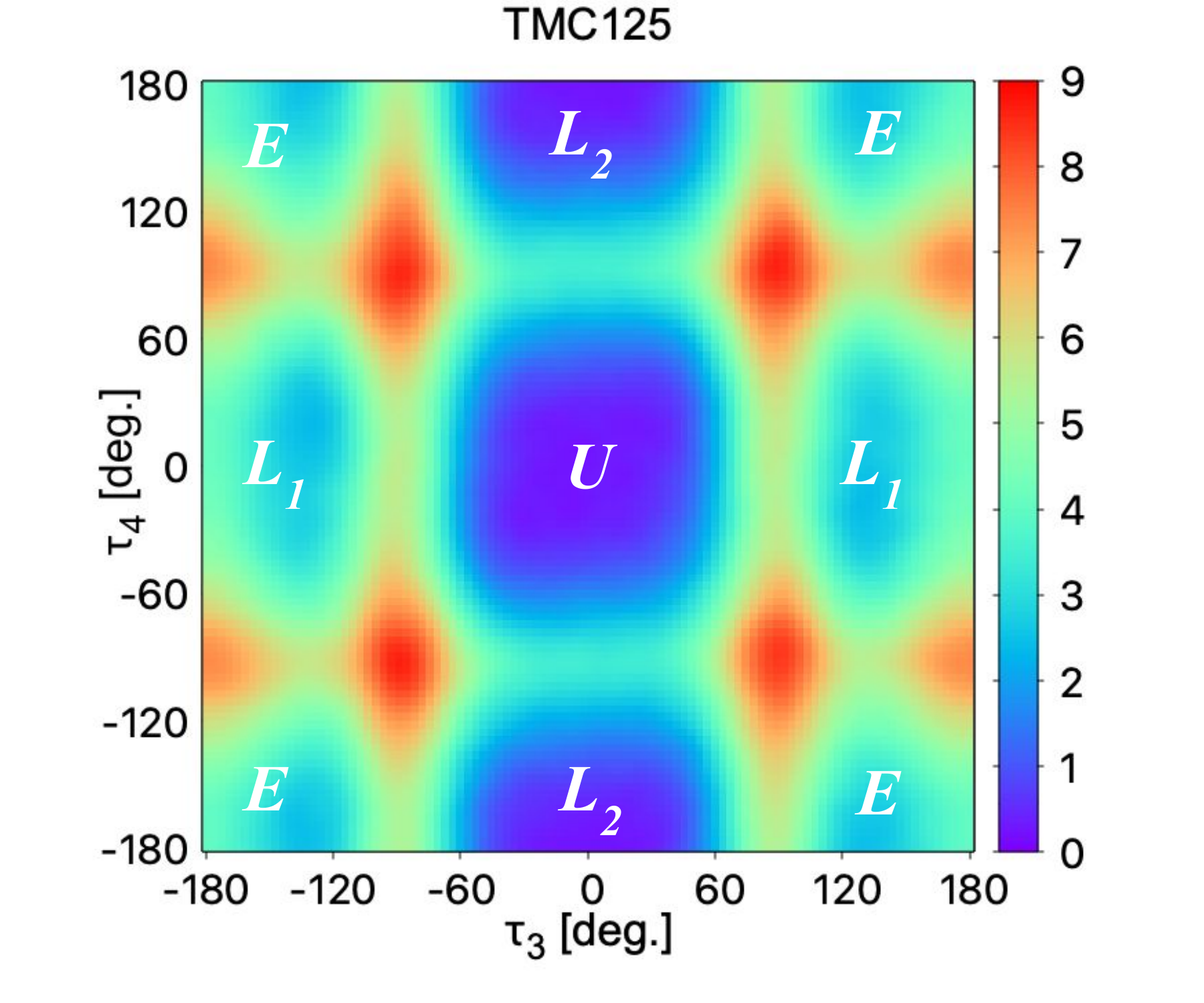}}

 \sidecaption{subfig:b1}
  \raisebox{-\height}{\includegraphics[width=0.70\textwidth]{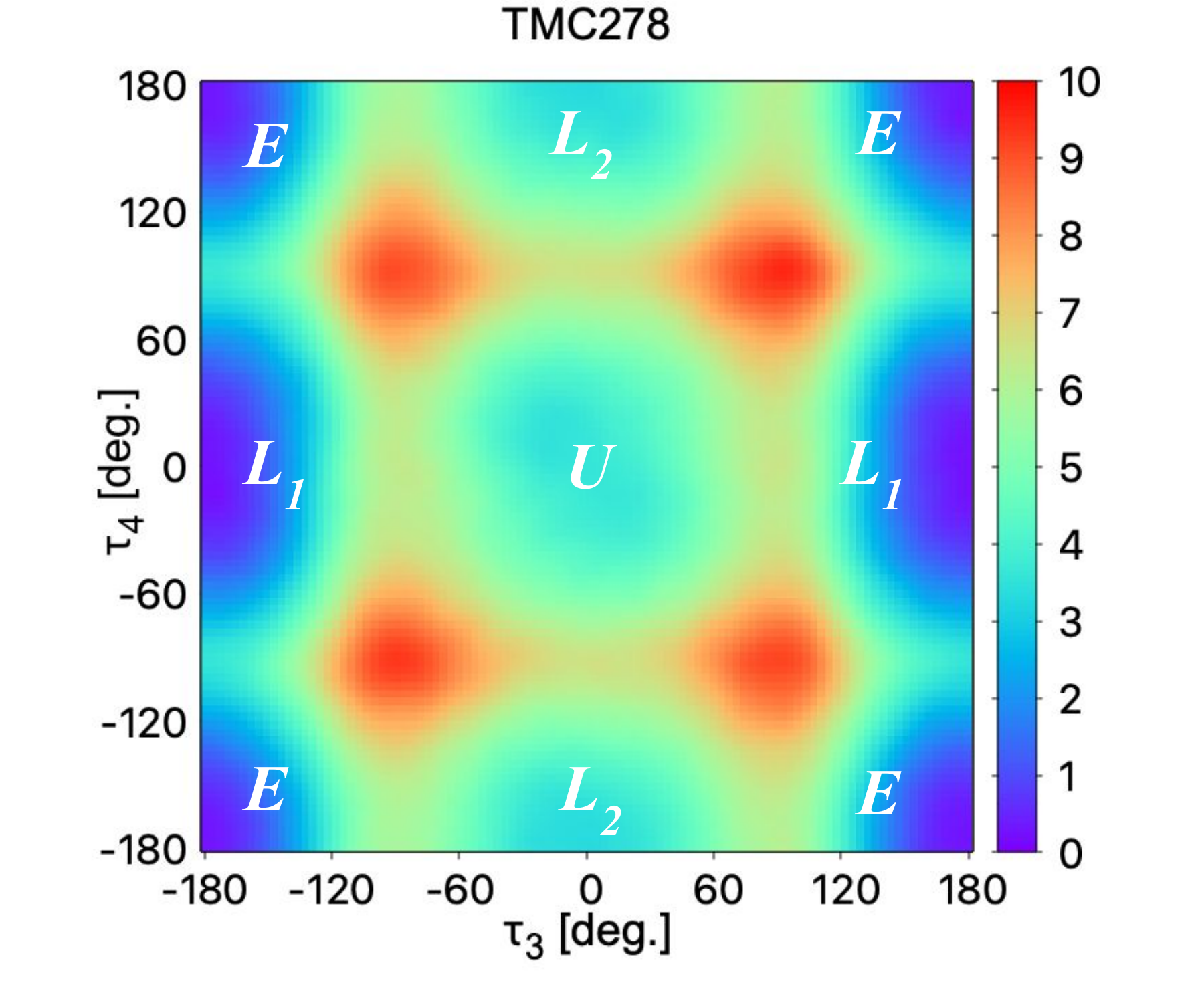}}%
  \hfill

  \caption{\label{fig:heatmap} The intramolecular potential of mean force with respect to the  $\tau_3$ and $\tau_4$ torsional angles of (a) TMC125 and (b) TMC278 in water. The potential of mean force is expressed in kcal/mol according to the color scale to the right of each plot. The labels of the conformational states ($U$ for U-shaped, $E$ for extended, and $L_1$ and $L_2$ for the two L-shaped conformations) follow the nomenclature in reference \citenum{Gallicchio2012b}. The $U$ state corresponds to the conformational state of the ligand bound to HIV-RT.}   
\end{figure}

As measured by the variance, metadynamics conformational sampling reduces the statistical fluctuations of the self-RBFEs in the case of TMC278 but increases it for TMC125 (Table \ref{tab:replicates-hiv}). As further discussed below, this is one of several examples we encountered where the more extensive conformational exploration afforded by metadynamics does not necessarily benefit ligands that do not significantly reorganize upon binding. The reorganization free energy estimates we obtained (Table \ref{tab:self-rbfe-reorg}) confirm that TMC278 suffers a much larger reorganization penalty than TMC125. Hence, accelerated conformational sampling helps reduce the statistical fluctuations of TMC278 by increasing the rate of interconversions between the $E$ and $L$ conformations predominant in solution (Figure \ref{fig:heatmap}) to the $U$ conformations required for binding. In contrast, TMC125 is already predominantly in the bound U-shaped conformation in solution (Figure \ref{fig:heatmap}), and exploring conformational states that do not contribute significantly to binding hurts convergence rather than enhances it. 

\begin{table}
\caption{\label{tab:replicates-hiv} Self-RBFE replicates for the complexes of HIV-RT with TMC125 and TMC278.} 
\centering
 \begin{tabular}{lcc}
 Replicate & \multicolumn{1}{c}{$\Delta\Delta G_b$(ATM)$^{a}$} & \multicolumn{1}{c}{$\Delta\Delta G_b$(ATM+MetaD)$^{a}$} \\ %[0.5ex] 
\hline
\multicolumn{3}{c}{ TMC125} \\
1 &	0.05	&	1.12	\\
2 &	-0.81	&	-0.25	\\
3 &	-0.33	&	1.80	\\
4 &	0.83	&	-0.64	\\
5 &	0.75	&	0.17	\\
\hline
%AUE$^{a}$  &  0.55 & 0.80 \\
average$^{a}$  & $0.10 \pm 0.62$ & $0.44 \pm 0.90$ \\
variance$^{b}$   & 0.49 & 1.01 \\

 \hline
\multicolumn{3}{c}{ TMC278} \\
1 &	1.80	&	0.03	\\
2 &	1.07	&	1.35	\\
3 &	1.06	&	0.44	\\
4 &	3.75	&	0.79	\\
5 &	-3.40	&	0.46 \\
\hline
%AUE$^{a}$  &  2.22 & 0.61 \\
average$^{a,c}$  & $0.86 \pm 2.34$ & $0.61 \pm 0.44$ \\
variance$^{b}$ & 6.88 & 0.24 \\
 \hline \\
\end{tabular}
\begin{flushleft}\small
$^a$In kcal/mol
$^b$In (kcal/mol)$^2$
$^c$Errors are reported as twice the standard error of the mean.
\end{flushleft}
\end{table}

\begin{figure*}
    \centering
    \includegraphics[scale=0.60]{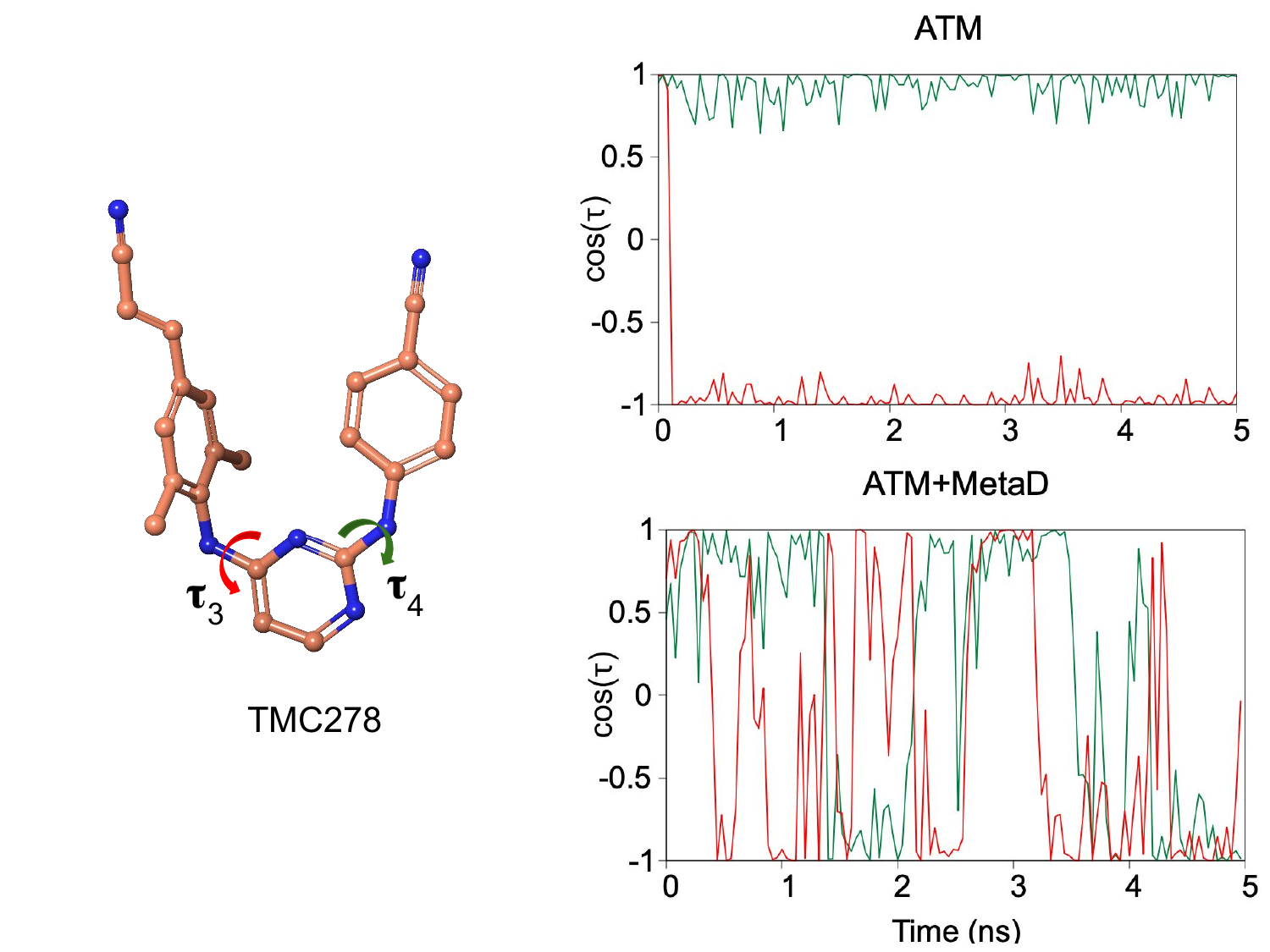} 
    \caption{\label{fig:time-traj-tmc278} Time trajectories with and without metadynamics sampling of the torsional angles $\tau_3$ (red) and $\tau_4$ (green) of one of the copies of TMC278 for a replica of the self-RBFE ATM Hamiltonian replica exchange simulation that is predominantly near the $\lambda = 0$ state when TMC278 is unbound.    }   
\end{figure*}

With this knowledge in mind, we tested whether metadynamics-based accelerated conformational sampling reduces the statistical uncertainty of the RBFE between TMC125 and TMC278 binding to HIV-RT. Since biasing the sampling of the dihedral angles of TMC125 proved ineffective, in these calculations we applied the metadynamics flattening potential only to TMC278 to better address its large conformational reorganization in solution. Indeed, as the results in Table \ref{tab:rbfe-hiv} show, metadynamics sampling yields an RBFE estimate with a much smaller statistical uncertainty (a variance of $0.16$ compared to $1.92$ (kcal/mol)$^2$ without metadynamics) at essentially the same computational cost. The reduction of the model variance, in this case, is due to the frequent transitions of TMC278 from solution to bound conformational states with metadynamics. With standard MD sampling, in contrast, TMC278 remains trapped in the extended solution conformations while in solution and is unable to equilibrate with the bound state of the complex (Figure  \ref{fig:time-traj-tmc278}).

Experimentally, TMC278 is a slightly better or equivalent inhibitor of HIV-RT than TMC125.\cite{tmc125tmc278} Without metadynamics sampling, ATM predicts that TMC125 is instead significantly more potent than TMC278 ($\Delta \Delta G_b = -1.49$ kcal/mol) albeit with a large uncertainty. With metadynamics sampling, the RBFE estimate is much closer to the expected value and with significantly smaller uncertainty. The overestimation of the potency of TMC125 relative to TMC278 without accelerated conformational sampling is attributed to the reduced ability of TMC278 to visit bound-competent conformations while in solution without the help of the metadynamics flattening potential.

\begin{table}[t]
\caption{\label{tab:rbfe-hiv} RBFE replicates for the TMC278-TMC125 ligand pair binding to the HIV-RT receptor} 
\centering
 \begin{tabular}{lcc}
 Replicate & \multicolumn{1}{c}{$\Delta\Delta G_b$(ATM)$^{a}$} & \multicolumn{1}{c}{$\Delta\Delta G_b$(ATM+MetaD)$^{a}$} \\ [0.5ex] 
\hline
\multicolumn{3}{c}{HIV-RT TMC278-TMC125} \\
1 &	-1.60	&	-0.34	\\
2 &	-2.33	&	-1.23	\\
3 &	-3.83	&	-0.26	\\
4 &	-0.12	&	-0.41	\\
5 &	-2.80	&	-0.78	\\
\hline
average$^{a,c}$  & $-1.49 \pm 1.24$ & $-0.40 \pm 0.18$ \\
variance$^{b}$  & $1.92$ &  $0.16$ \\
\hline \\
\end{tabular}
\begin{flushleft}\small
$^a$In kcal/mol.
$^b$In (kcal/mol)$^2$
$^c$Errors are reported as twice the standard error of the mean.
\end{flushleft}
\end{table}

The results of the self-RBFE bias and variance analysis for the benchmark sets from Schindler et al.\ are shown in Table \ref{tab:self-rbfe-reorg}. The full list of self-RBFE estimates for each ligand in the set is provided in the Supplementary Information. To compare with the results of Chen et al.\cite{chen2023performance}, these calculations did not employ metadynamics accelerated conformational sampling. The model bias measured as the average of the self-RBFEs is within statistical uncertainty for all three sets. However, the statistical spread of the estimates as measured by the average unsigned errors (AUEs) is relatively large ($0.87$, $0.85$, and $0.80$ kcal/mol 
for the c-Met, Syk, and CDK8 sets, respectively, see the Supplementary Information) and comparable to the AUEs relative to the experiments of the RBFEs between dissimilar ligands of the same sets obtained by Chen at al.\cite{chen2023performance} ($0.98$, $1.13$, and $1.50$ kcal/mol, respectively) using the same setup and force field employed here.

\begin{figure}
  \centering
  \sidecaption{subfig:a2}
  \raisebox{-\height}{\includegraphics[width=0.45\textwidth]{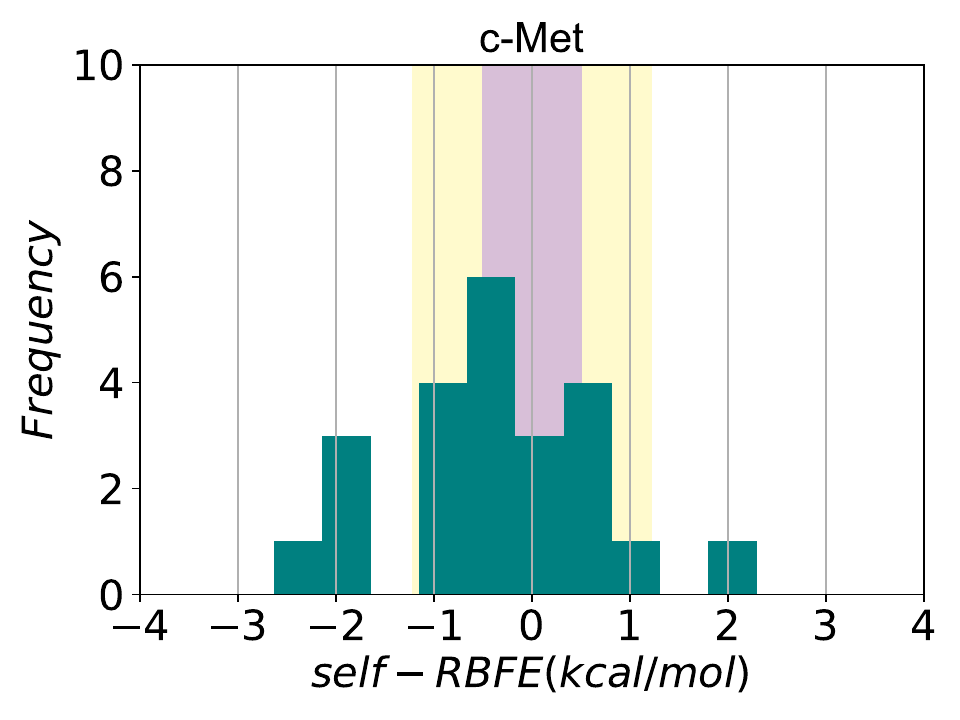}}

 \sidecaption{subfig:b2}
  \raisebox{-\height}{\includegraphics[width=0.45\textwidth]{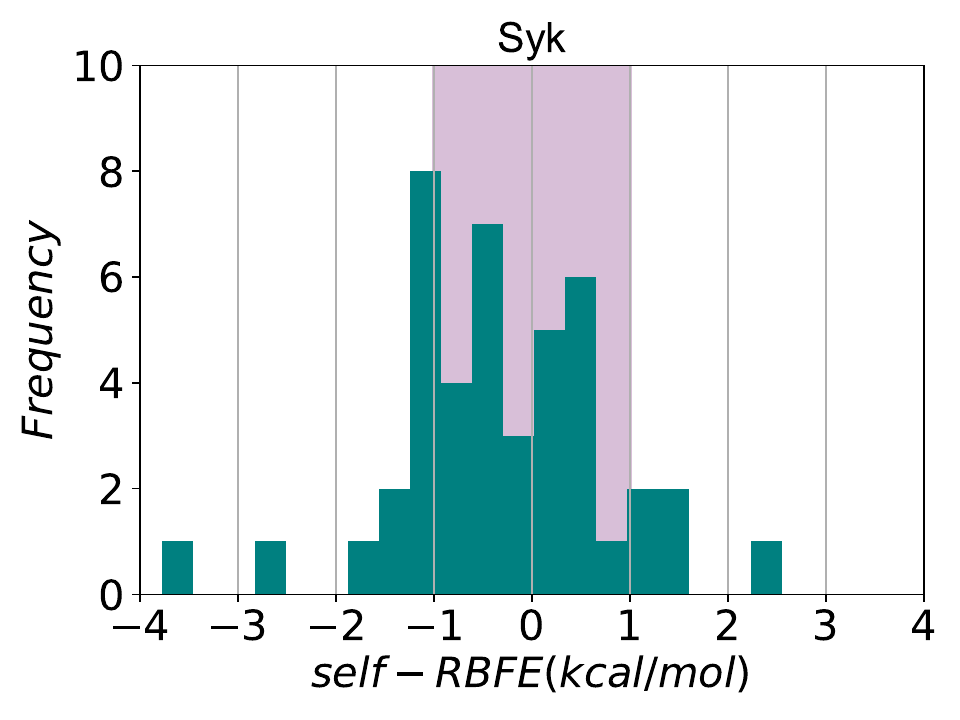}}%
  \hfill
 \sidecaption{subfig:c2}
  \raisebox{-\height}{\includegraphics[width=0.45\textwidth]{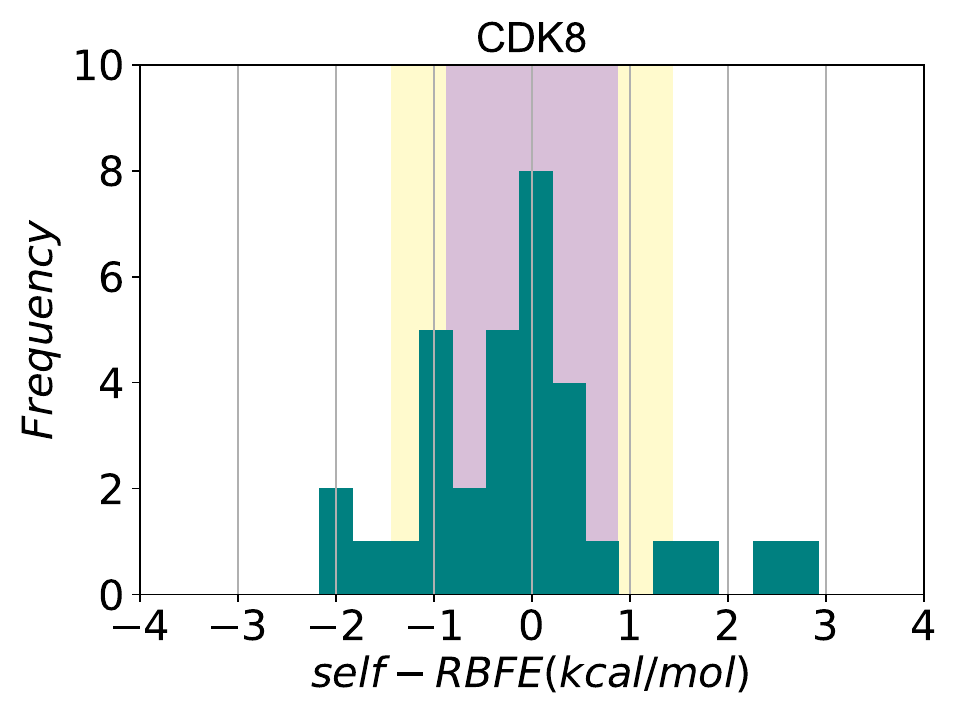}}
  \caption{\label{fig:histograms} Distributions of self-RBFE values ($\Delta\Delta G_b^{ATM}$) for the for the sets (a) c-Met (24 complexes) (b) Syk (44 complexes), and (c) CDK8 (32 complexes). The yellow and purple shaded regions in the plots represent the standard deviation of the self-RBFE estimates from replicates of a subset of complexes obtained with (purple) and without (yellow) metadynamics sampling (see Supplementary Information tables 4, 5, and 6). The standard deviation range is centered at zero and shown at the one sigma level ($\pm \sigma$). For the Syk set, the difference of the standard deviations with and without metadynamics is too small to be distinguishable.}
\end{figure}

\begin{figure*}
    \centering
    \includegraphics[scale=0.80]{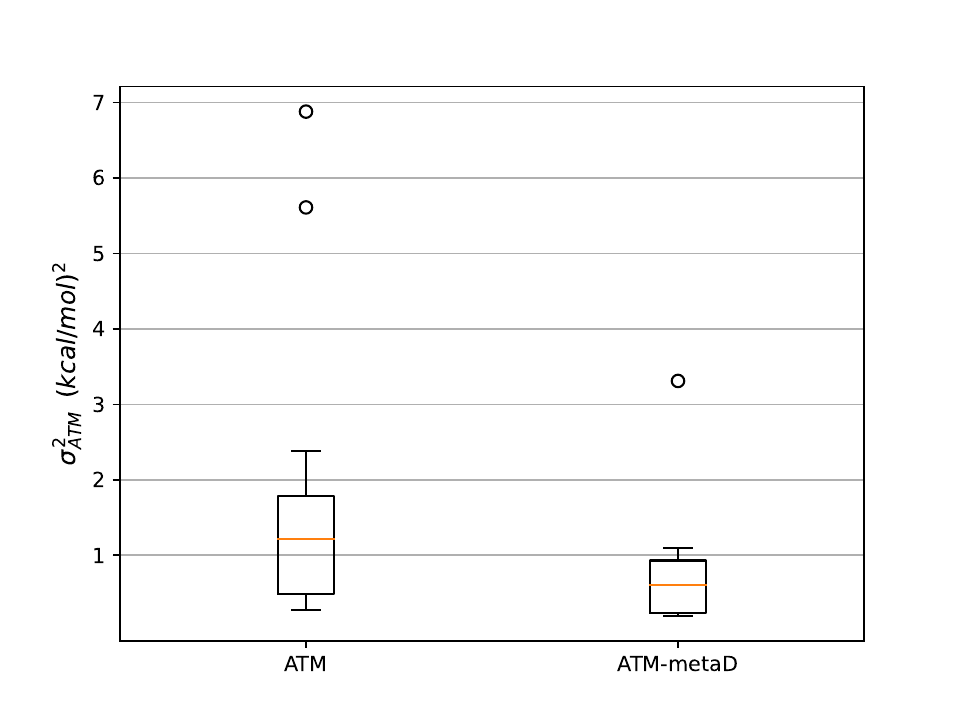} 
    \caption{\label{fig:aue} Boxplots of the distributions of ATM self-RBFE model variances with and without metadynamics accelerated conformational sampling. Metadynamics sampling of the ligands affords smaller statistical fluctuations on average with fewer outliers.}
\end{figure*}

To gain a better understanding of the sources of statistical noise affecting the self-RBFE estimates, we performed a similar analysis as for the TMC125 and TMC278 inhibitors of HIV-RT above on a small subset of randomly picked ligands of the c-Met, Syk, and CDK8 sets. We conducted five self-RBFE replicates for each complex in the subset to estimate the model bias and variance with and without metadynamics accelerated conformational sampling (Supplementary tables 4, 5 and 6). The results show that self-RBFE ATM's model bias is consistently small and within statistical uncertainty (Table \ref{tab:self-rbfe-reorg}). They also show that the self-RBFEs without accelerated conformational sampling vary considerably from one ligand to another. We observe that metadynamics sampling reduces the model variance in most cases (Figure \ref{fig:aue}). Without metadynamics sampling, the median self-RBFE model variance ATM calculations is approximately $1.2$ (kcal/mol)$^2$ with some outliers reaching up to $7$ (kcal/mol)$^2$. In contrast, the median self-RBFE variance with ATM-metaD is significantly reduced to approximately $0.5$ (kcal/mol)$^2$ with fewer and less severe outliers (Figure \ref{fig:aue}). 

\begin{table}[t]
\caption{\label{tab:self-rbfe-reorg}  Values of the average, variance, and ligand reorganization free energy of the HIV-RT set and the subset of c-Met, Syk and CDK8 complexes. } 
\centering
\sisetup{separate-uncertainty}
\begin{tabular}{l S[table-format = 3.2(2)] c S[table-format = 3.2(2)] c  S[table-format = 3.2(2)]}
    & \multicolumn{2}{c}{ ATM} &  \multicolumn{2}{c}{ ATM+MetaD} & \\
\multicolumn{1}{l}{Ligand}  & \multicolumn{1}{c}{average$^{a,c}$} & \multicolumn{1}{c}{variance$^{b}$} & \multicolumn{1}{c}{average$^{a,c}$} & \multicolumn{1}{c}{variance$^{b}$} & \multicolumn{1}{c}{$\Delta G_{\rm reorg}$$^{a,c}$}\\ [0.5ex] 
\hline
\multicolumn{6}{c}{ HIV-RT} \\
TMC125   &  0.10(62)   & $0.49$  &  0.44(90) & $1.01$  & 0.37(70) \\
TMC278   &  0.86(234)  & $6.88$  &  0.61(44) & $0.24$ & 3.97(60) \\

\multicolumn{6}{c}{ c-Met} \\
CHEMBL3402742 &  0.01(98)   & $1.22$ & -0.03(44) & $0.24$ & 0.76(90) \\
CHEMBL3402749 &  1.31(64)   & $0.53$ & -0.41(42) & $0.21$ & 2.65(80) \\
CHEMBL3402755 & -0.39(104)  & $1.38$ &  0.17(40) & $0.20$ & 2.74(76) \\
  
 \multicolumn{6}{c}{ Syk } \\
CHEMBL3264999 & -0.01(120) & $1.79$  & -0.20(160) & $3.31$ &  3.45(90) \\
CHEMBL3265004 &  0.16(78)  & $0.75$  & -0.14(56)  & $0.40$ &  1.50(80) \\
CHEMBL3265034 &  0.59(106) & $1.42$  &  0.14(58)  & $0.42$ &  1.90(83) \\
CHEMBL3265037 & -0.21(46)  & $0.27$  &  0.14(70)  & $0.61$ &  1.80(80) \\

\multicolumn{6}{c}{ CDK8 } \\
Ligand 17 &  0.13(138) & $2.38$ &  -0.15(74) & $0.67$ & 3.35(76) \\
Ligand 19 &  0.96(212) & $5.61$ &  -0.07(84) & $0.93$ & 3.80(74) \\
Ligand 37 & -0.21(48)  & $0.28$ &   0.03(94) & $1.09$ & 1.23(82) \\
Ligand 38 & -0.21(46)  & $0.28$ &   0.14(84) & $0.91$ & 0.42(80) \\
\hline
\end{tabular}
\begin{flushleft}\small
$^a$In kcal/mol.
$^b$In (kcal/mol)$^2$
$^c$Errors are reported as twice the standard error of the mean.
\end{flushleft}
\end{table}

\begin{figure*}
    \centering
    \includegraphics[scale=0.90]{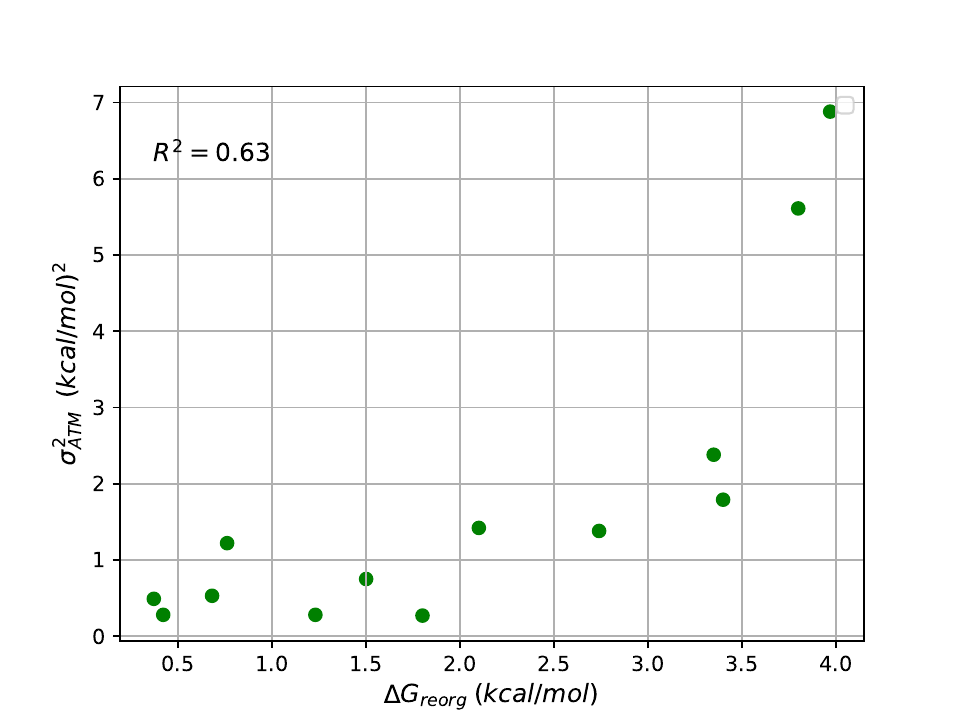} 
    \caption{\label{fig:correlation} Scatterplot of ATM's self-RBFE model variance without metadynamics conformational sampling vs.\ the ligand reorganization free energy. Ligands that undergo strong conformational reorganization upon binding tend to display greater self-RBFE statistical fluctuations. }
\end{figure*}

Following the insights we obtained for the HIV-RT systems, we set out to explain the large variations in self-RBFE statistical fluctuations by testing the hypothesis that self-RBFE model variance is larger for ligands that undergo a large reorganization upon binding and that, consequently, metadynamics accelerated conformational sampling would be most beneficial for ligands suffering a large reorganization free energy penalty but less so for ligands that do not appreciably reorganize upon binding. Accordingly, we calculated the reorganization free energy for binding for the same subset of ligands (Table \ref{tab:self-rbfe-reorg}) and found a strong statistical correlation between the self-RBFE model variance and the magnitude of the ligand reorganization free energy (Figure \ref{fig:correlation}). This analysis reveals that about 63\% of the spread of model variance is explained by the reorganization of the ligand from the solution to the bound state. 

\begin{figure*}
    \centering
    \includegraphics[scale=0.60]{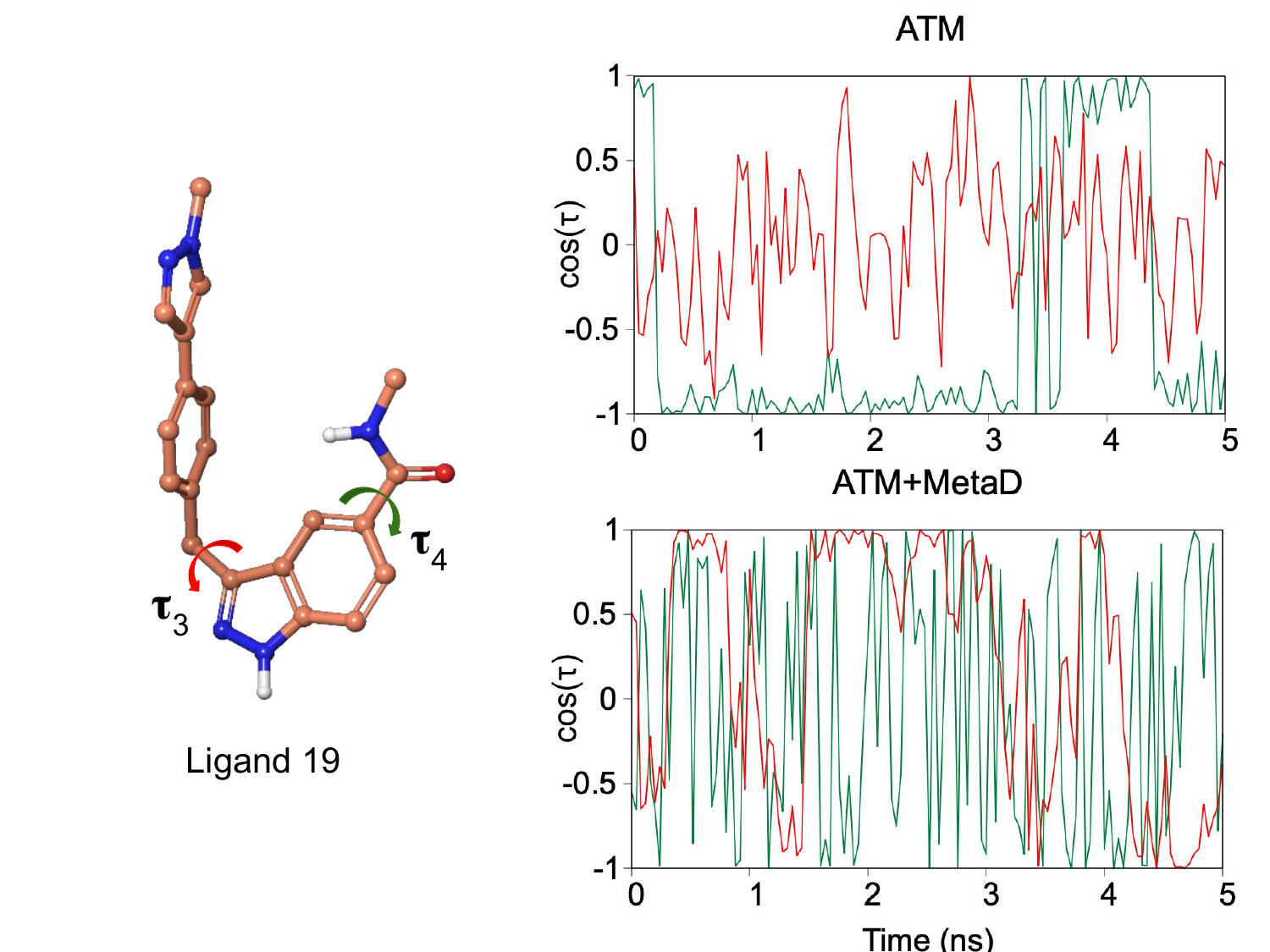} 
    \caption{\label{fig:time-traj-19} Time trajectories with and without metadynamics sampling of the torsional angles $\tau_3$ (red) and $\tau_4$ (green) of one of the copies of ligand 19 of the CDK8 benchmark set for a replica of the self-RBFE ATM Hamiltonian replica exchange simulation that is predominantly near the $\lambda = 0$ state when ligand 19 is unbound. }   
\end{figure*}

The result provides support to the conclusion that, generally, accelerated sampling of the internal degrees of the ligands reduces the variance by better representing the equilibration between the unbound and bound states of the ligand. One representative case of the many we observed is self-RBFE calculation of ligand 19 of the CDK8 protein target (Figure \ref{fig:time-traj-19}). Similarly to the TMC278 in Figure \ref{fig:time-traj-tmc278} above, we actively sampled the $\tau_3$ and $\tau_4$ torsional angles of 19 using metadynamics. Without metadynamics sampling (Figure \ref{fig:time-traj-19}, left panel), the ligand often gets trapped in conformations favored in solution. For example, the hindered amide torsional angle represented by $\tau_4$ visits predominantly in-plane configurations (Figure \ref{fig:time-traj-19}, left panel, green trace). In contrast, with metadynamics sampling (ATM-MetaD), the ligand visits a wide range of conformations that span preferred bound and solution conformations, resulting in smaller self-RBFE statistical fluctuations  (Table {\ref{tab:self-rbfe-reorg}}).

\section{Discussion}

Binding free energy models are routinely validated against experimental datasets.\cite{wang2015accurate,schindler2020large,kuhn2020assessment,gapsys2020large,ganguly2022amber,hahn2022bestpractices,sabanes2023validation,chen2023performance} However because the models' biases -- the differences between the binding free energies and the values that the models would yield after infinite sampling -- are not known, it is unclear how the results of these benchmarks can be used to improve the accuracy of the models. It is often found, for example, that increasing the level of accuracy of the force field does not significantly change the average deviation from the experimental data,\cite{gapsys2022pre,hahn2023current} raising the possibility that statistical uncertainties, rather than inaccuracy of the model, dominate the observed model errors. One largely uncontrollable source of noise is the target experimental activity data, which varies depending on the conditions and does not always reliably report the actual binding strength.\cite{Brown2009a,tresadern2022impact} However, a likely more pervasive source of statistical noise is the random variation of the free energy models' predictions due to incomplete conformational sampling\cite{mobley2012let,procacci2019solvation} during the relatively short simulations that are the norm in applied work. Thus, to truly appreciate the performance of binding free energy models, it is critical to investigate not only the models' biases but also their variances.

In this work, we studied the statistical properties of relative binding free energy calculations for a ligand relative to itself (self-RBFE) to assess the bias and variance of the Alchemical Transfer Method (ATM) for systems relevant to structure-based drug discovery. Because the true value of a self-RBFE is known to be zero, we were able to estimate the bias and the variance independently by running a series of replicate calculations of the same length. We found that, while ATM's self-RBFE bias is generally small, its variance can be occasionally quite large. We found, in fact, that the magnitudes of the self-RBFE statistical errors we measured here account for $50$ to $90$\% the average ATM errors of recent RBFE benchmarks on the same systems.\cite{chen2023performance} This result is even more remarkable, considering that self-RBFEs are more straightforward transformations and likely suffer less bias than RBFEs between dissimilar ligands. Overall, the results obtained here raise the tantalizing possibility that a significant reduction of ATM's model error can be achieved by better conformational sampling alone. 

To begin along this path, we found that, for ligands that reorganize upon binding, large statistical fluctuations are caused by conformational trapping of the ligands in their solution conformations, thereby preventing proper equilibration between unbound and bound states. As a further confirmation, we established that accelerated metadynamics conformational sampling along torsional degrees of the ligand can significantly reduce the statistical variance of self-RBFEs, and, in one case (TMC125 vs.\ TMC278), it yielded an RBFE estimate in closer agreement with the experiment.

The reorganization free energy of binding,\cite{Yang2009,Gallicchio2011adv,Foloppe2016TowardsUT,Foloppe2021TheRE} also known in the literature as induced-fit and intramolecular strain,\cite{Kar2010,cournia2020rigorous} has a profound effect on binding affinities.\cite{DeLorbe2009,Gallicchio2012b,wickstrom2016parameterization,Wickstrom2013,zheng2017conformational,pinheiro2023magic} It originates from the free energy cost for the ligand and the receptor\cite{fajer2023quantitatively} to assume binding-competent conformations from their conformational ensembles when free in solution. Thus, it is important to capture reorganization effects in alchemical binding free energy calculations. Often, this requires accelerated conformational sampling algorithms to avoid conformational trapping.\cite{wang2013modeling} In this work, we accelerated the sampling of the intramolecular torsional degrees of the ligand by employing a biasing potential derived from metadynamics\cite{bussi2018metadynamics} that flattens the conformational landscape of the ligand in solution. This approach significantly enhanced the convergence of binding free energy estimates for ligands that reorganize upon binding. However, the more thorough exploration of the degrees of freedom of the ligand unnecessarily expanded the conformational sampling of ligands that do not reorganize upon binding, resulting in a slowdown rather than an enhancement of their rate of convergence. This approach also does not directly address the equally important reorganization of the receptor.\cite{fajer2023quantitatively} Nevertheless, here accelerated metadynamics conformational sampling had a significant net positive effect in general, mostly by removing large outliers.

\section{Conclusions}

We evaluated the self-RBFE technique to assess the intrinsic statistical fluctuations of the Alchemical Transfer Method (ATM) relative binding free energy estimator. We illustrate the approach to a small set of HIV-RT inhibitors and apply it to large datasets often used to benchmark the accuracy of alchemical relative binding free energy methods. We thoroughly examined the variance and bias of the self-RBFE calculations to gain insights into the source of errors often observed when comparing calculated RBFEs to experimental values. We find evidence that a significant fraction of these errors could be due to statistical noise. Hence, the reduction of statistical fluctuations should be taken as a priority when attempting to use the results of benchmarking studies to improve the accuracy of free energy models. Self-RBFE tests are limited to the validation of dual- and hybrid-topology alchemical relative binding free energy methods. However, the results obtained here confirm the benefits of assessing free energy models in general by means of validation tests on transformations with known true values, such as cycle-closure,\cite{Mey2020Best} to assess the inherent statistical fluctuations of free energy estimators.

Notably, we find that ligand reorganization is a significant contributing factor to the statistical variance of binding free energy estimates and that accelerated conformational sampling of the degrees of freedom of the ligand can drastically reduce the time to convergence.  In this work, we employ a metadynamics-based approach to enhance the sampling of slow torsional degrees of freedom of the ligand that often cause conformational trapping. 

The additional computational cost of self-RBFE tests and metadynamics-based conformational sampling and analysis of ligand conformational variability in solution is relatively minor compared to the already high demands of relative binding free energy campaigns and the substantial benefits coming from the greater reliability and the deeper assessment of the predictions. We recommend wide adoption of these practices.

\section{Acknowledgements}

We acknowledge support from the National Science Foundation (NSF CAREER 1750511 to E.G.).

\section{Supplementary Information}

The Supplementary Information document lists the self-RBFEs of the $110$ complexes from the Schindler et al.\ benchmark set, and the results of the replicates of the 11 complexes in the subset analyzed with and without accelerated metadynamics sampling.

\section{Data and Software Availability}

The AToM-OpenMM software used in this study is publicly available at {\tt https://github.com/\-Gallicchio-Lab/\-AToM-OpenMM}. 
The molecular system files, AToM-OpenMM input files, and the analysis scripts for each ligand in the benchmark set used here are publicly available on GitHub at {\tt https://github.com/\-sheenam1509/\-self-rbfe.git}.

%\bibliographystyle{unsrt}
%do not change main.bib, add new references to add2main.bib
%\bibliography{main,add2main}
\providecommand{\latin}[1]{#1}
\makeatletter
\providecommand{\doi}
  {\begingroup\let\do\@makeother\dospecials
  \catcode`\{=1 \catcode`\}=2 \doi@aux}
\providecommand{\doi@aux}[1]{\endgroup\texttt{#1}}
\makeatother
\providecommand*\mcitethebibliography{\thebibliography}
\csname @ifundefined\endcsname{endmcitethebibliography}
  {\let\endmcitethebibliography\endthebibliography}{}

\clearpage
\noindent {\bf ToC Graphic}\vspace{1cm}

\includegraphics{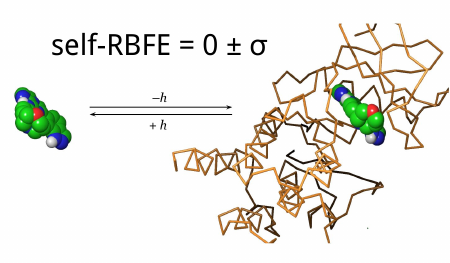} 

\clearpage
\noindent {\large\bf Supplementary Information}

This supplementary document contains the self-RBFE values for the entire set of c-Met, Syk, and CDK8 complexes discussed in the paper. Additionally, self-RBFEs from the five replicates for the subset of complexes with ATM and ATM+MetaD chosen from c-Met, Syk, and CDK8 are listed. 

\begin{supptable}
\centering
\caption{Self-RBFE estimates for the c-Met complexes.}
\label{tab:cmet-self-rbfe}
{\scriptsize
\begin{tabular}{lcc}
Ligand       & $\Delta\Delta G_b$(ATM)$^{a}$  \\ \hline
CHEMBL3402741	&	 0.78 \\
CHEMBL3402742   &	-1.80 \\
CHEMBL3402743	    &	 0.43 \\
CHEMBL3402744	    &  -0.32 \\
CHEMBL3402745    &	-0.57 \\
CHEMBL3402747   &	-0.97 \\
CHEMBL3402748   &  2.29 \\
CHEMBL3402749	    &	0.39 \\
CHEMBL3402750    &	-0.91 \\
CHEMBL3402751   &	-0.77 \\
CHEMBL3402752	&	0.09 \\
CHEMBL3402753	&	0.37 \\
CHEMBL3402755	&	-2.12 \\
CHEMBL3402756	&	-0.82 \\
CHEMBL3402757	&	-0.17 \\
CHEMBL3402758	&	-0.28 \\
CHEMBL3402759	&	-0.33 \\
CHEMBL3402760	&	0.20 \\
CHEMBL3402761	&	0.02 \\
CHEMBL3402762	&	1.21 \\
CHEMBL3402763	&	-0.55 \\
CHEMBL3402764	&	-2.08 \\
CHEMBL3402765	&	-2.63 \\
\hline
AUE$^{a}$  &  0.87 \\
average$^{a,c}$  &  $-0.37 \pm 0.46$  \\
variance$^{b}$ &  $1.25$  \\
\hline
\end{tabular}
\begin{flushleft}\small
$^a$In kcal/mol.
$^b$In (kcal/mol)$^2$
$^c$Errors are reported as twice the standard error of mean.
\end{flushleft}
}
\end{supptable}

\begin{supptable}
\caption{ Self-RBFE estimates for the Syk complexes.}
\centering
\label{tab:syk-self-rbfe}
{\scriptsize
\begin{tabular}{lcc}
Ligand       & $\Delta\Delta G_b$(ATM)$^{a}$ \\ \hline
CHEMBL3259820	&	-0.95 \\
CHEMBL3264994	&	-1.22 \\
CHEMBL3264995	&	-0.93 \\
CHEMBL3264996	&	0.97 \\
CHEMBL3264997	&	-0.48 \\
CHEMBL3264998	&	-1.26 \\
CHEMBL3264999	&	1.17 \\
CHEMBL3265000	&	0.29 \\
CHEMBL3265001	&	-0.02 \\
CHEMBL3265002 &	-1.49 \\
CHEMBL3265003	&	0.41 \\
CHEMBL3265004	&	0.37 \\
CHEMBL3265005	&	0.16 \\
CHEMBL3265006	&	-2.75 \\
CHEMBL3265008	&	0.51 \\
CHEMBL3265009	&	-1.07 \\
CHEMBL3265010	&	-3.77 \\
CHEMBL3265011	&	-0.05 \\
CHEMBL3265012	&	-0.51 \\
CHEMBL3265013	&	-0.64 \\
CHEMBL3265014	&	-0.05 \\
CHEMBL3265015	&	-0.79 \\
CHEMBL3265016	&	-0.96 \\
CHEMBL3265017	&	0.03 \\
CHEMBL3265018	&	1.39 \\
CHEMBL3265019	&	-1.71 \\
CHEMBL3265020 &	-0.42 \\
CHEMBL326502	&	-1.15 \\
CHEMBL3265022	&	0.15 \\
CHEMBL3265023	&	0.56 \\
CHEMBL3265024	&	-0.43 \\
CHEMBL3265025	&	-0.64 \\
CHEMBL3265026	&	-0.42 \\
CHEMBL3265027	&	-1.19 \\
CHEMBL3265028	&	0.37 \\
CHEMBL3265029	&	-0.34 \\
CHEMBL3265030	&	0.59 \\
CHEMBL3265031	&	-1.05 \\
CHEMBL3265032	&	1.40 \\
CHEMBL3265033	&	-0.66 \\
CHEMBL3265034	&	2.55 \\
CHEMBL3265035	&	1.25 \\
CHEMBL3265036	&	-0.57 \\
CHEMBL3265037	&	0.21 \\
\hline
AUE$^{a}$  &  0.86 \\
average$^{a,c}$  & $-0.30 \pm 0.34$  \\
variance$^{b}$  &  $1.23$ \\
\hline
\end{tabular}
\begin{flushleft}\small
$^a$In kcal/mol.
$^b$In (kcal/mol)$^2$
$^c$Errors are reported as twice the standard error of the mean.
\end{flushleft}
}
\end{supptable}

\begin{supptable}
\caption{Self-RBFE estimates for the CDK8 complexes.}
\centering
\label{tab:cdk8-self-rbfe}
{\scriptsize
\begin{tabular}{lcc}
\hline
Complex Pair       & $\Delta\Delta G_b$(ATM)$^{a}$  \\ \hline
13	&	0.01	\\
14	&	0.16	\\
15	&	-1.32	\\
16	&	-0.27	\\
17	&	-2.17	\\
18	&	0.03	\\
19	&	2.93	\\
20	&	-0.65	\\
21	&	0.85	\\
22	&	1.75	\\
23	&	0.44	\\
24	&	-2.03	\\
25	&	-0.87	\\
26	&	0.48	\\
27	&	0.50	\\
28	&	-0.16	\\
29	&	0.19	\\
30	&	-0.38	\\
31	&	-0.11	\\
32	&	2.42	\\
33	&	-0.14	\\
34	&	-0.93	\\
35	&	0.14	\\
36	&	-1.05	\\
37	&	-0.97	\\
38	&	-0.95	\\
39	&	1.25	\\
40	&	0.25	\\
41	&	-0.79	\\
42	&	0.12	\\
43	&	-0.37	\\
44	&	-1.55	\\
45	&	0.15	\\			
\hline
AUE$^{a}$  &  $0.80$ \\
average$^{a,c}$  &  $-0.09 \pm 0.38$ \\
variance$^{b}$  & $1.23$ \\
\hline
\end{tabular}
\begin{flushleft}\small
$^a$In kcal/mol.
$^b$In (kcal/mol)$^2$
$^c$Errors are reported as twice the standard error of the mean.
\end{flushleft}
}
\end{supptable}

\begin{supptable}
\caption{\label{tab:replicates-cmet}  Self-RBFE replicates for a subset of the c-Met complexes.} 
\centering
{\scriptsize
 \begin{tabular}{lcc}
 Replicate & \multicolumn{1}{c}{$\Delta\Delta G_b$(ATM)$^{a}$} & \multicolumn{1}{c}{$\Delta\Delta G_b$(ATM+MetaD)$^{a}$} \\ [0.5ex] 
\hline
\multicolumn{3}{c}{ CHEMBL3402742} \\
1 & -1.80	& -0.79 \\ 
2 & 0.12 & 0.21  \\ 
3 & -0.04	&	0.19  \\ 
4 & 1.03	&	-0.21  \\ 
5 & 0.76	&	0.47  \\ 
\hline
average$^{a,c}$  &  $0.01 \pm 0.98$ & $-0.03 \pm 0.44$  \\
variance$^{b}$  &  $1.22$ & $0.24$ \\
 \hline
\multicolumn{3}{c}{CHEMBL3402749} \\
1 & 0.39	&	-0.10 \\ 
2 & 1.49	&	0.20  \\ 
3 & 2.39	&	-0.79  \\ 
4 & 1.07	&	-0.90  \\ 
5 & 1.20	&	-0.45  \\ 
\hline
average$^{a,c}$  & $1.31 \pm 0.64$  & $-0.41 \pm 0.42$ \\
variance$^{b}$  & $0.53$ &  $0.21$ \\
\hline
\multicolumn{3}{c}{CHEMBL3402755} \\
1 & -2.12	&	0.52 \\ 
2 & 0.18	&	0.35  \\ 
3 & 0.77	&	-0.07  \\ 
4 & 0.91	&	0.54  \\ 
5 & -0.96	&	-0.49  \\ 
\hline
average$^{a,c}$  & $-0.39 \pm 1.04$ &  $0.17 \pm 0.40$ \\
variance$^{b}$  & $1.38$ & $0.20$ \\
\hline
overall variance$^{b}$ & $2.04$ & $0.77$ \\
\hline
\end{tabular}
\begin{flushleft}\small
$^a$In kcal/mol.
$^b$In (kcal/mol)$^2$
$^c$Errors are reported as twice the standard error of the mean.
\end{flushleft}
}
\end{supptable}

\begin{supptable}
\caption{\label{tab:replicates-syk} Self-RBFE replicates for a subset of the Syk complexes.}
\centering
{\scriptsize
\begin{tabular}{lcc}
 Replicate & \multicolumn{1}{c}{$\Delta\Delta G_b$(ATM)} & \multicolumn{1}{c}{$\Delta\Delta G_b$(ATM+MetaD)} \\ [0.5ex] 
\hline
\multicolumn{3}{c}{CHEMBL3264999} \\

1 &	1.17	&	2.15	\\
2 &	-1.79	&	-0.32	\\
3 &	0.93	&	0.83	\\
4 &	-1.09	&	-2.65	\\
5 &	0.71	&	-1.00	\\ 
\hline
average$^{a,c}$  & $-0.01 \pm 1.20$ &  $-0.20 \pm 1.60$  \\
variance$^{b}$  & $1.79$ &  $3.31$ \\
\hline \\
\multicolumn{3}{c}{CHEMBL3265004 } \\

1 &	0.37	&	-0.46	\\
2 &	-0.91	&	0.33	\\
3 &	-0.26	&	-0.81	\\
4 &	0.15	&	-0.48	\\
5 &	1.44	&	0.70	\\
\hline
average$^{a,c}$  &  $0.16 \pm 0.78$ & $-0.14 \pm 0.56$ \\
variance$^{b}$  &  $0.75$ &  $0.40$ \\
\hline\\
\multicolumn{3}{c}{CHEMBL3265034} \\

1 &	2.55	&	0.56	\\
2 &	-0.16	&	0.77	\\
3 &	0.77	&	-0.90	\\
4 &	-0.47	&	0.26	\\
5 &	0.26	&	-0.01	\\
\hline
average$^{a,c}$  & $0.59 \pm 1.06$ &  $0.14 \pm 0.58$ \\
variance$^{b}$  &  $1.42$ & $0.42$ \\
\hline  \\
\multicolumn{3}{c}{CHEMBL3265037} \\

1 &	0.21	&	0.59	\\
2 &	-1.00	&	-1.00	\\
3 &	-0.47	&	1.10	\\
4 &	0.21	&	0.02	\\
5 &	0.02	&	-0.01	\\
\hline
average$^{a,c}$  &  $-0.21 \pm 0.46$ & $0.14 \pm 0.70$ \\
variance$^{b}$  & $0.27$ & $0.61$ \\
\hline  
overall variance$^{b}$ & $0.98$ & $1.02$ \\
\hline
\end{tabular}
\begin{flushleft}\small
$^a$In kcal/mol.
$^b$In (kcal/mol)$^2$
$^c$Errors are reported as twice the standard error of the mean.
\end{flushleft}
}
\end{supptable}

\begin{supptable}
\caption{\label{tab:replicates-cdk8}  Self-RBFE replicates for a subset of the CDK8 complexes.}
\centering
{\scriptsize
\begin{tabular}{lcc}
 Replicate & \multicolumn{1}{c}{$\Delta\Delta G_b$(ATM)} & \multicolumn{1}{c}{$\Delta\Delta G_b$(ATM+MetaD)} \\ [0.5ex] 
\hline
\multicolumn{3}{c}{Ligand 17} \\

1 &	-2.17	&	0.72	\\
2 &	1.75	&	0.68	\\
3 &	0.55	&	-0.61	\\
4 &	-0.58	&	-1.11	\\
5 &	1.09	&	-0.44	\\
\hline
average$^{a,c}$  &  $0.13  \pm 1.38$ &  $-0.15 \pm 0.74$  \\
variance$^{b}$  & $2.38$ & $0.67$  \\
\hline \\
\multicolumn{3}{c}{Ligand 19 } \\

1 &	2.93	&	1.18	\\
2 &	1.58	&	-0.66	\\
3 &	-1.39	&	0.21	\\
4 &	3.34	&	0.29	\\
5 &	-1.67	&	-1.34	\\
\hline
average$^{a,c}$  & $0.96 \pm 2.12$ & $-0.07 \pm 0.84$ \\
variance$^{b}$  & $5.61$ & $0.93$  \\
\hline \\
\multicolumn{3}{c}{Ligand 37} \\

1 &	$-0.97$	&	$1.52$	\\
2 &	$0.26$	&	$0.55$	\\
3 &	$0.34$	&	$-0.65$	\\
4 &	$-0.26$	&	$-1.15$	\\
5 &	$-0.40$	&	$-0.13$	\\
\hline
average$^{a,c}$  &  $-0.21 \pm 0.48$  &  $0.03 \pm 0.94$ \\
variance$^{b}$  & $0.28$ &  $1.09$  \\
\hline \\
\multicolumn{3}{c}{Ligand 38} \\

1 &	$-0.95$	 &	$1.27$	\\
2 &	$0.50$	 &	$-0.52$	\\
3 &	$-0.14$  &	$0.81$	\\
4 &	$-0.13$	 &	$-1.08$	\\
5 &	$-0.32$	 &	$0.20$	\\
\hline
average$^{a,c}$  & $-0.21 \pm 0.46$ &  $0.14 \pm 0.84$ \\
variance$^{b}$   & $0.28$ & $0.91$  \\
\hline 
overall variance$^{b}$ & $2.04$ & $0.77$ \\
\hline
\end{tabular}
\begin{flushleft}\small
$^a$In kcal/mol.
$^b$In (kcal/mol)$^2$
$^c$Errors are reported as twice the standard error of the mean.
\end{flushleft}
}
\end{supptable}

\end{document}